\documentclass[
  aps,
  pre,
  reprint,
  amsmath,
  amssymb,
  superscriptaddress,
  nofootinbib
]{revtex4-2}
\usepackage[T1]{fontenc}
\usepackage[utf8]{inputenc}
\usepackage{graphicx}   
\usepackage{pmboxdraw}  
\usepackage[most]{tcolorbox}    	
\usepackage{hyperref}
\usepackage[capitalize]{cleveref}
\crefname{section}{Sec.}{Secs.}
\Crefname{section}{Sec.}{Secs.}
\crefname{appendix}{Appendix}{Appendices}
\Crefname{appendix}{Appendix}{Appendices}

\tcbuselibrary{skins,breakable}

\usetikzlibrary{arrows.meta,calc}
\definecolor{sA}{HTML}{EFB366}   
\definecolor{sB}{HTML}{7FA9D6}   
\definecolor{sC}{HTML}{93C08A}   
\newcommand{\sweepsites}[1][]{%
  \node[site,tA,#1] (s1) at (0.75,2.05) {1};
  \node[site,tB,#1] (s2) at (1.72,1.72) {2};
  \node[site,tC,#1] (s3) at (0.42,0.95) {3};
  \node[site,tA,#1] (s4) at (2.45,1.05) {4};
  \node[site,tA,#1] (s5) at (3.05,2.35) {5};
  \node[site,tC,#1] (s6) at (1.88,2.52) {6};
}
\newcommand{\sweepbonded}[2]{%
  \node[site,#1] (s7) at (0.55,0.18) {7};
  \node[site,#2] (s8) at (1.62,0.30) {8};
}

\begin{document}

\title{Simulation and Network Assembly Pipelines for \protect\\ Dynamically Bonded Soft Materials}

\author{Tanner A. Wilcoxson}
\affiliation{McKetta Department of Chemical Engineering, University of Texas at Austin, Austin, Texas 78712, USA}

\author{Tianhao Li}
\affiliation{Department of Chemistry, New York University, New York, New York 10003, USA}
\affiliation{Simons Center for Computational Physical Chemistry, New York University, New York, New York 10003, USA}

\author{Tyla R. Holoman}
\affiliation{McKetta Department of Chemical Engineering, University of Texas at Austin, Austin, Texas 78712, USA}

\author{Gaurav Mitra}
\affiliation{Department of Chemistry, New York University, New York, New York 10003, USA}
\affiliation{Present Address: Department of Biomedical Engineering and Center for Biomolecular Condensates, Washington University in St. Louis, St. Louis, MO 63130, USA}

\author{Thomas M. Truskett}
\email{truskett@umich.edu}
\affiliation{Department of Chemical Engineering and Biointerfaces Institute, University of Michigan, Ann Arbor, Michigan 48109, USA}
\affiliation{McKetta Department of Chemical Engineering, University of Texas at Austin, Austin, Texas 78712, USA}

\author{Glen M. Hocky}
\email{hockyg@nyu.edu}
\affiliation{Department of Chemistry, New York University, New York, New York 10003, USA}
\affiliation{Simons Center for Computational Physical Chemistry, New York University, New York, New York 10003, USA}

\date{\today}

\begin{abstract}
Soft materials linked by reversible covalent or supramolecular bonds form a diverse class of assemblies with promising applications from nanoscience to medicine.
Experiments on such systems typically probe bulk assembly phase behavior and rheological properties, but it remains difficult to measure how microscopic bonding kinetics and the mechanical properties of the constituent elements give rise to the observed bulk properties. 
Coarse-grained molecular dynamics simulations offer a route to develop physical, mechanistic insights that bridge microscopic and macroscopic behaviors, but most simulation approaches do not provide a way to control individual bond kinetics. 
Moreover, many approaches introduced to address this limitation were developed for bespoke applications and do not readily generalize to a wide range of self-assembling systems.
In this work, we present pySNAP (Simulation and Network Assembly Pipelines), a modular open source Python-based software platform that integrates tunable dynamic bonding with a workflow, template, and analysis setup that lets users study a wide range of systems with only small changes to input files. 
This platform is built on the HOOMD-blue MD engine, which provides fast GPU-accelerated molecular dynamics simulations, and integrates DyBond, a dynamic bonding plugin that forms and breaks bonds consistent with an equilibrium distribution. 
DyBond supports bonding between multiple types of partner species and is GPU-accelerated. 
Around this core, the \texttt{snap\_simulate} package
compiles a directory of setup parameter files into a HOOMD-blue
simulation, and the \texttt{snap\_workflow} package orchestrates the resulting
parameter sweeps across workstations and high-performance computing schedulers. 
In this article, we describe the theory behind simulated dynamic bonding and how to use the package, from setting up a parameter sweep to analyzing its results. 
We demonstrate the framework across a diverse range of dynamically bonded systems, showing that its modular design can accommodate distinct interaction mechanisms, geometries, and physical scenarios within a unified simulation workflow, while enabling both straightforward reproduction of existing models and rapid construction of more complex composite systems.
\end{abstract}

\maketitle

\section{Introduction}
Soft materials often comprise molecular or colloidal building blocks that assemble through reversible dynamic bonds, leading to emergent structure and macroscopic properties~\cite{zhang2013general,kay2016dynamic,bomboi2016re, richardson2019,marro2022constitutionally,peng2020,song2020programmable,dominguez2020assembly,kang2022colorimetric,kang2025colloidal,kim2025dense}.
Equilibrium behavior and dynamic structure formation in these materials reflect the choice of building blocks and the kinetics of bonding interactions. The latter can depend sensitively on environmental conditions such as solvent choice, pH, temperature, or salt concentration, and can produce functionalities such as self-healing, stimuli-responsiveness, reprocessability, and recyclability~\cite{Bertsch2023Self-HealingRegeneration,wang2020self} desired for applications in 3D printing, tissue engineering, and drug delivery, to mention a few~\cite{rosales2016design,loebel2017shear}.

Building blocks in these materials span size scales from small organic molecules, nanoparticles, polymers, proteins, nucleic acids, and DNA-origami tiles to micron-sized colloids. 
When reversible `dissociative' bonds, which continually form and break, drive assembly into networks, the equilibrium fraction of bonds can be readily tuned by changing the aforementioned environmental factors, which modify binding and dissociation separately. In an alternative `associative' type of dynamic chemistry, formed bonds are not broken, only swapped between neighboring partners during dynamic network rearrangement, while conserving the total number of bonds during the exchange. 
Whether associative or dissociative, the soft materials formed by these chemistries are collectively coined covalent adaptable networks (CANs), with associative polymeric networks commonly referred to as vitrimers~\cite{Kloxin2013CovalentSystems, Webber2022DynamicInteractions,Karatrantos2024MolecularReview}.

While molecular dynamics (MD) simulations based on ab initio or machine-learned interatomic potentials allow chemists to model the reactions that control the properties of these soft materials~\cite{sun2023modeling,akhtar2026molecular}, the mismatch between the timescale of the reaction kinetics and that of large-scale network rearrangements means that these approaches are generally inappropriate for studying assembly driven by dynamic bonding.
In contrast, coarse-grained MD simulations can model the dynamic assembly of many nano- or micron-scale components, but incorporating even phenomenological models for dynamic covalent bonding in standard MD software is nontrivial. 
The demand to model and understand these systems has driven development of hybrid computational approaches to simulate networks formed through dynamic bonding~\cite{Ciarella2018DynamicsRelaxation,Gissinger2017ModelingSimulations,Vashisth2018AcceleratedPolymers,Blanco2024TheMethod,Rao2024APolymers,mitra2023coarse,Jedlinska2024EffectsCoacervates,Freedman2017ANetworks,Lugo2023ASystems,Yan2022TowardAssemblies,Popov2016MEDYAN:Networks,Liu2024AEquilibrium}; for a recent review, see Ref.~\cite{holoman2026simulating}.

Here, we adopt the dynamic bonding model previously implemented in Refs.~\cite{mitra2023coarse,shu2024mesoscale,holoman2026coarse}, which is well suited for probing the effects of changing both binding and unbinding rates on assembly processes in coarse-grained MD simulations and for describing soft materials under kinetic or thermodynamic control. This physics is particularly relevant for describing the structural evolution of nanoscale colloidal networks assembled via bifunctional linkers~\cite{dominguez2020assembly,Sherman2023,ofosu2026universal,kang2022colorimetric,kang2023modular,kang2023structural,xu2023dynamic, kang2025colloidal,hallstrom2026decoding,kwon2026connectivity} or associating networks of multivalent organic molecules such as tetraPEG~\cite{FitzSimons2020PreferentialAdditions,fitzsimons2022effect,crowell2023shear,crowell2026leveraging,holoman2026simulating}. With minor modifications, the framework is also suitable for studying polymerization through reversible association~\cite{Oyarzun2018EfficientNanoparticles,zhang2021sequence}, phase separation of multivalent proteins~\cite{shu2024mesoscale,dai2026modular}, assembly of colloids with mobile linkers~\cite{angioletti2014mobile,bachmann2016melting,mcmullen2018freely,mitra2023coarse,judd2025statistical}, and more.

Dynamic bonding chemistry is essential to network designability. It is common to functionalize two types of building blocks with complementary bonding pairs such as a thiol and a conjugate acceptor (thia-Michael addition) or an aldehyde and a hydrazide. These pairs form strong yet labile bonds, with high bond-pair specificity and orthogonality (i.e., different pairs do not cross-react). Formation and dissociation rates of such dynamic bonds can be tuned by pH, temperature, and molecular
substitution on the bonding partner~\cite{scheutz2019adaptable,reuther2019tunable,cudjoe2018strong,Zhang2022Structure-Reactivity-PropertyNetworks,
FitzSimons2020PreferentialAdditions}. Mutual orthogonality is essential for programmable assembly~\cite{reuther2019tunable}, but some assembly strategies also require competing reactions. Thus, simulation frameworks must be able to treat several chemically distinct or even competitive reactions at once,
rather than just a single bonded pair. For example, a linker and a capping molecule competing for the same site are, in the model, two reactions drawing on one pool of reactive sites.

Compelling design questions about assembly follow from this control. For example, which combinations of building block/linker composition and bond strength produce an equilibrium gel rather than an arrested, nonequilibrium network, and by what pathway? How do the topological defects that the network
carries---loops closed within a single building block, dangling ends, or
double links between the same pair of neighbors---depend on the linker
stoichiometry, and how much elasticity do they cost? How do the elastic moduli
and the stress relaxation time follow from the bond kinetics, given that the two
can be varied independently at fixed bond free energy? 

Answering any of these means mapping a large parameter space rather than running
a single simulation. The independent variables include the building-block
functionality $f$, the linker-to-cap and linker-to-colloid ratios, the packing fraction,
and the strength and kinetics of the bond, and the observables of interest are
often defined only after the network has formed and then been driven. This
combinatorial cost, rather than the cost of any individual trajectory, is the
practical obstacle to using computation to guide experiment.

To address these challenges, we present an end-to-end computational toolbox for simulating reversible macromolecular assemblies, built on two distinct but complementary layers: the DyBond simulation engine and the pySNAP workflow framework. At the simulation layer, DyBond is a dedicated C++/CUDA plugin for HOOMD-blue~\cite{Anderson2020HOOMD-blue:Simulations} designed to introduce dynamic covalent and supramolecular chemistry into MD. DyBond evaluates bond creation and cleavage during runtime using a Metropolis Monte Carlo scheme tailored for independent candidate pairs, enabling efficient GPU-accelerated modeling of arbitrary reactive particle types (see Sec.~\ref{sec:theory}).

To orchestrate the complex multi-stage pipelines and extensive parameter sweeps required to investigate these systems, we surround DyBond with pySNAP (Simulation and Network Assembly Pipelines). pySNAP decouples simulation configuration and execution into two lightweight Python packages: \texttt{snap\_simulate} provides declarative building and runtime automation for HOOMD-blue, handling macromolecular mixture generation (such as telechelic and patchy polymers), automated box compression, and automated autocorrelation-based runtime termination to prevent wasted compute; \texttt{snap\_workflow} provides a general-purpose parametric execution framework that manages inter-stage dependencies, parameter sweeps, and dynamic resource packing across workstations and Slurm-managed HPC clusters. By decoupling the dynamic bonding physics (DyBond) from the high-throughput simulation orchestration (pySNAP), this architecture provides both specialized physical fidelity and a scalable, automated pipeline for soft-matter exploration.

\section{DyBond-v2: Theory and Implementation}
\label{sec:theory}

Addition and removal of bonds is implemented in HOOMD-blue in the DyBond-v2 plugin, which stochastically changes HOOMD-blue's bond table at discrete intervals during an MD run. The scheme generalizes the dynamic bonding model of Ref.~\cite{mitra2023coarse}, which in turn was adapted from a plugin developed to study epoxies~\cite{Thomas2018RoutineDynamics}. Here we describe the model as currently implemented, emphasizing details of the three features that are new: support for an arbitrary number of chemically distinct reactions competing for shared reactive sites, an acceptance criterion evaluated with whichever bond potential the integrator uses, and a GPU implementation whose stochastic decisions are independent of thread scheduling. The competition between reactions is validated against an exactly computed answer in \cref{sec:validation}.  The DyBond-v2 repository can be found at \url{https://github.com/hocky-research-group/DyBond-v2}. 

\subsection{Bonding as a two-state reaction}
\label{sec:rates}
For each distinct dynamic bonding reaction, $\alpha$, a subset of the particles in the simulation are designated reactive sites consisting of the reactants of $\alpha$, $i$ and $j$. 
The bonded and unbonded states are connected by rate constants $k^\alpha_\mathrm{on}$ and $k^\alpha_\mathrm{off}$. Because these sites are explicit particles carried by the molecular building blocks, the valence of a macromolecule is set by the number of reactive sites it carries rather than by a separate parameter.
Hence, in our scheme, each individual particle may only participate in zero or one dynamic bond.

Treating each $ij$ pair within a distance of $[r_\mathrm{min},r_\mathrm{max}]$ from each other as a two-state system, the equilibrium constant and the corresponding bond free energy are
\begin{equation}
\label{eq:epsilon}
K^\alpha = \frac{k^\alpha_\mathrm{on}}{k^\alpha_\mathrm{off}}, \qquad
\varepsilon^\alpha \equiv -\frac{\Delta G^\alpha}{k_\mathrm{B}T}=\ln K^\alpha.
\end{equation}
Separating the two rate constants is crucial, because the same equilibrium network can be reached quickly or slowly through different dynamical pathways, and also the system can reach certain structures through kinetic control that are not the minimum free energy.

Every $n$ integration steps of size $\Delta t$, a bond updater performs one sweep of attempted bond formation and bond breaking for each $\alpha$.
During each sweep, the probabilities are calculated as,
\begin{equation}
\label{eq:probs}
P^{\alpha,0}_\mathrm{on} = k^\alpha_\mathrm{on}\,n\,\Delta t, \qquad
P^{\alpha,0}_\mathrm{off} = k^\alpha_\mathrm{off}\,n\,\Delta t,
\end{equation}
which yield physically consistent rates so long as both probabilities remain below unity. Thus the maximum accessible rate constant of all reactions is determined by $k_\mathrm{max} = 1/(n\,\Delta t)$, at which point reactions are only governed by the availability of partners dictated by the local density.

\subsection{Acceptance criterion}
\label{sec:metropolis}

Creating a bond between two particles at separation $r$ adds a bond energy
$U^\alpha_\mathrm{b}(r)$ to the system that was not there before. Accepting every
proposal with probability $P^{\alpha,0}_\mathrm{on}$ would therefore inject energy and drive the system away from the equilibrium implied by \cref{eq:epsilon}.
There are multiple schemes for modifying these rates to maintain detailed balance~\cite{Marbach2023Coarse-grainedLinkers}, and following Ref.~\cite{mitra2023coarse} we chose to add a Metropolis factor~\cite{metropolis1953equation} to the binding probability, while the breaking
probability remains unchanged,
\begin{equation}
\label{eq:pon}
P^\alpha_\mathrm{on}(r) = P^{\alpha,0}_\mathrm{on}\,
\exp\!\left[-\frac{U^\alpha_\mathrm{b}(r)}{k_\mathrm{B}T}\right], \qquad
P^\alpha_\mathrm{off} = P^{\alpha,0}_\mathrm{off}.
\end{equation}
where $T$ is the instantaneous kinetic temperature computed by a HOOMD-blue compute instance each time the updater triggers.
Therefore, all of the energetic dependence sits in the on-rate and the ratio of
forward to reverse probabilities for a pair at fixed separation reproduces the
Boltzmann weight of the bond.
Putting the energy dependence in the forward step suppresses the formation of strained bonds, preventing the scheme from injecting large forces that the MD thermostat must resolve.

For a harmonic bond of stiffness $k_\mathrm{b}$ and rest length $r_0$, the binding probability in \cref{eq:pon} reduces to the form used in Ref.~\cite{mitra2023coarse},
\begin{equation}
\label{eq:pon_harmonic}
P^\alpha_\mathrm{on}(r) = P^{\alpha,0}_\mathrm{on}\,
\exp\!\left[-\frac{k_\mathrm{b}(r - r_0)^2}{2k_\mathrm{B}T}\right].
\end{equation}
In our revised implementation of DyBond, however, $U_\mathrm{b}(r)$ is not assumed harmonic. 
Rather, DyBond-v2 uses HOOMD-blue's bond energy evaluator passed during each reaction's initialization.
This implementation requires setting bond parameters only when initializing HOOMD's Bond objects, preventing mismatch between MD bond forces and DyBond's Metropolis energy calculations.
HOOMD-blue implements Harmonic, FENEWCA, and ``Tether'' bonds, as well as the ability to use arbitrary tabulated potentials;
of these, the Harmonic, FENEWCA, and ``Tether'' evaluators are currently accepted by DyBond.
Additionally, DyBond utilizes a templated C++ architecture parameterized by the bond evaluator class, facilitating straightforward integration with other custom HOOMD-blue bond potentials.

\subsection{The bonding sweep}
\label{sec:sweep}
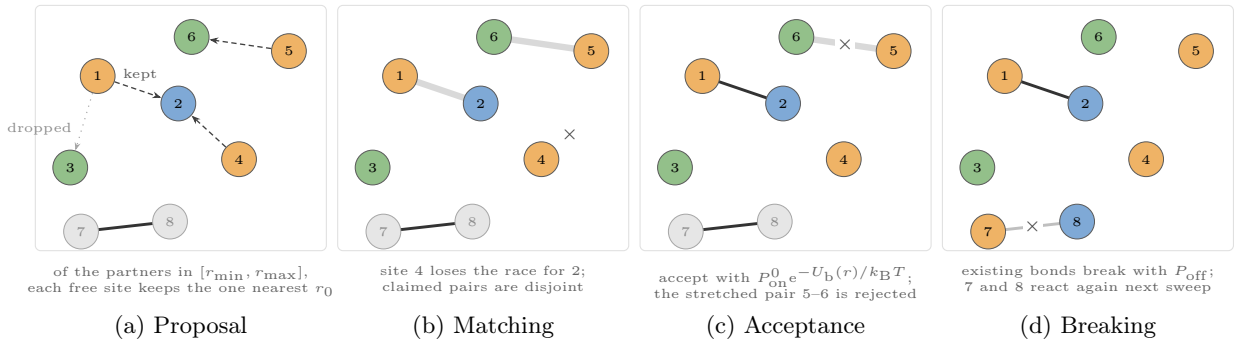
\begin{figure*}[t]
\centering
\begin{tikzpicture}[
  x=1.10cm, y=1.10cm,
  site/.style={circle, draw=black!65, line width=0.35pt, minimum size=4.6mm,
               inner sep=0pt, font=\tiny},
  tA/.style={fill=sA}, tB/.style={fill=sB}, tC/.style={fill=sC},
  busy/.style={fill=black!10, draw=black!35, text=black!45},
  prop/.style={-{Stealth[length=3.4pt]}, dashed,
               dash pattern=on 2pt off 1.3pt, line width=0.5pt, black!75},
  drop/.style={-{Stealth[length=3pt]}, dotted, line width=0.45pt, black!35},
  claim/.style={line width=2.4pt, black!15, line cap=round},
  bond/.style={line width=1.1pt, black!80},
  win/.style={line width=1.1pt, black!80},
  cross/.style={font=\scriptsize, text=black!75, inner sep=0pt},
  plabel/.style={font=\small, anchor=north, text depth=0pt},
  note/.style={font=\tiny, black!62, anchor=north, align=center},
  panel/.style={draw=black!12, line width=0.4pt, rounded corners=2pt}
]
\def\panelw{3.5}  \def\panelh{2.9}

\begin{scope}[xshift=0cm]
  \draw[panel] (0,-0.05) rectangle (\panelw,\panelh);
  \draw[bond] (0.55,0.18) -- (1.62,0.30);
  \sweepbonded{busy}{busy}
  \sweepsites
  \draw[drop] (s1) -- (s3)
    node[pos=0.62, left=0.5pt, font=\tiny, black!45] {dropped};
  \draw[prop] (s1) -- (s2)
    node[pos=0.55, above=0.5pt, font=\tiny, black!65] {kept};
  \draw[prop] (s4) -- (s2);
  \draw[prop] (s5) -- (s6);
  \node[note] at (1.75,-0.13) {of the partners in
                               $[r_\mathrm{min},r_\mathrm{max}]$,\\
                               each free site keeps the one nearest $r_0$};
  \node[plabel] at (1.75,-0.72) {(a) Proposal};
\end{scope}

\begin{scope}[xshift=4.00cm]
  \draw[panel] (0,-0.05) rectangle (\panelw,\panelh);
  \draw[bond] (0.55,0.18) -- (1.62,0.30);
  \draw[claim] (0.75,2.05) -- (1.72,1.72);
  \draw[claim] (3.05,2.35) -- (1.88,2.52);
  \sweepbonded{busy}{busy}
  \sweepsites
  \node[cross] at ($(2.45,1.05)+(0.34,0.30)$) {$\times$};
  \node[note] at (1.75,-0.13) {site 4 loses the race for 2;\\
                               claimed pairs are disjoint};
  \node[plabel] at (1.75,-0.72) {(b) Matching};
\end{scope}

\begin{scope}[xshift=8.00cm]
  \draw[panel] (0,-0.05) rectangle (\panelw,\panelh);
  \draw[bond] (0.55,0.18) -- (1.62,0.30);
  \draw[win] (0.75,2.05) -- (1.72,1.72);
  \draw[claim] (3.05,2.35) -- (1.88,2.52);
  \sweepbonded{busy}{busy}
  \sweepsites
  \node[cross, fill=white, inner sep=0.6pt]
    at ($(3.05,2.35)!0.5!(1.88,2.52)$) {$\times$};
  \node[note] at (1.75,-0.13) {accept with
      $P^{0}_\mathrm{on}\mathrm{e}^{-U_\mathrm{b}(r)/k_\mathrm{B}T}$;\\
      the stretched pair 5--6 is rejected};
  \node[plabel] at (1.75,-0.72) {(c) Acceptance};
\end{scope}

\begin{scope}[xshift=12.00cm]
  \draw[panel] (0,-0.05) rectangle (\panelw,\panelh);
  \draw[bond, black!25] (0.55,0.18) -- (1.62,0.30);
  \draw[win] (0.75,2.05) -- (1.72,1.72);
  \sweepbonded{tA}{tB}
  \sweepsites
  \node[cross, fill=white, inner sep=0.6pt]
    at ($(0.55,0.18)!0.5!(1.62,0.30)$) {$\times$};
  \node[note] at (1.75,-0.13) {existing bonds break with $P_\mathrm{off}$;\\
                               7 and 8 react again next sweep};
  \node[plabel] at (1.75,-0.72) {(d) Breaking};
\end{scope}
\end{tikzpicture}
\caption{One bonding sweep, drawn for a single reaction pair set at frozen
particle positions; the coordinates are identical in all four panels because a
sweep is evaluated between integration steps. Circles are reactive sites,
colored by particle type, and heavy lines are dynamic bonds. (a)~Every
unoccupied site scans the neighbor list and keeps the one free, type-compatible
partner lying in $[r_\mathrm{min},r_\mathrm{max}]$ whose separation is closest
to the bond rest length; site~1 accordingly proposes 2 (dashed) rather than the
more distant 3 (dotted), and sites 7 and 8, which already carry a bond, are
invisible to the search. (b)~Claims are resolved so that no site appears in more
than one proposal: sites 1 and 4 both want 2, and whichever claim lands first
wins, leaving the other site unpaired for this sweep. (c)~Each surviving pair is
accepted with the probability of \cref{eq:pon}; the stretched pair 5--6 carries a
large $U_\mathrm{b}(r)$ and is rejected, releasing its claim. (d)~Only then are
the existing bonds of this reaction tested for breaking. Because the proposal and
matching phases read the occupancies as they stood before this reaction was
processed, the pair 7--8 broken here cannot be re-formed by the same reaction
within this sweep, which is the independence condition that makes its trials
factorize. With several reactions registered, sites 7 and 8 remain available to
reactions processed later in the same sweep.}
\label{fig:sweep}
\end{figure*}
Bonds are added and removed by a sweep performed each time the updater is called,
built so that the individual trials within it are independent and the
correct equilibrium therefore follows from the criterion applied to each trial.
Every reactive site carries a flag recording whether it currently holds a
dynamic bond, and a site so marked is invisible to the bonding sweep, as candidate and
as partner alike; this flag is what enforces the unit valence of
\cref{sec:rates}.

For each reaction, $\alpha$, the sweep first determines which non-bonded particles should form bonds followed by which pre-existing bonds should break, with the former being executed in three phases and the latter in just one phase (\cref{fig:sweep}). In the bonding \emph{proposal} phase, every free site scans its entries in HOOMD-blue's neighbor list---whose cutoff is ensured to be at least $r^\alpha_\mathrm{max}$---and keeps one free, type-compatible, non-self partner whose separation lies within $[r^\alpha_\mathrm{min}, r^\alpha_\mathrm{max}]$: by default the suitable partner minimizing the distance to the bond rest length $r^\alpha_0$, $\left| r - r^\alpha_0 \right|$, where $r^\alpha_0$ is read from the bond potential attached to the reaction. In the \emph{matching} phase, proposals with overlapping reaction site candidates are resolved by claiming independent pairs. In the \emph{acceptance} phase, the surviving proposals are accepted with probability $P^\alpha_\mathrm{on}(r)$ as defined in \cref{eq:pon}; a rejected pair releases its claim. Pre-existing bonds corresponding to the current $\alpha$ are then tested for breaking with $P^\alpha_\mathrm{off}$. The updater then continues to the next $\alpha$; only after all reactions have been processed are the accumulated topology changes applied to the bond table.

Two properties of this ordering matter for correctness. First, the proposal and matching phases of a reaction read occupancies as they stood before that reaction was processed, and its existing bonds are tested for breaking only afterwards, so a bond broken by a reaction cannot re-form under that specific reaction. Second, matching guarantees that the trials within a single reaction sweep act on disjoint pairs of sites. Together these make the individual events independent, so that the total change in bonded pairs factorizes and detailed balance for the whole sweep follows from the criterion applied to each trial~\cite{mitra2023coarse}. What is given up is the ability of a pair to unbind and immediately rebind, an event whose omission matters only if $n\Delta t$ is long compared with the bond lifetime---the same condition, $k_\mathrm{off} n \Delta t \ll 1$, under which \cref{eq:probs} is a faithful representation of the reactions.

An arbitrary number of chemically distinct reactions may be registered with a single updater, and all of them act on the same configuration in every sweep. This generalizes Ref.~\cite{mitra2023coarse}, in which a reaction connected one complementary pair of binder types. Because the occupancy flag belongs to the site rather than to the reaction, reactions sharing a particle type draw on a common pool of reactive sites: a site claimed or bonded by one reaction is unavailable to the rest, so that the branching between competing products is governed by their relative rates. To keep that competition from favoring whichever reaction was registered first, the order in which the reactions are processed is redrawn at random each sweep. Because the reactions are processed one after another, a site released by the breaking phase of one reaction is free to form a bond in subsequent reactions processed in the same sweep, so a site may exchange between two competing chemistries within a single sweep even though no reaction re-forms a bond it has itself just broken. That competing reactions reach the correct equilibrium populations is checked exactly in a frozen four-reaction geometry in \cref{sec:validation}.

\subsection{Rate modifiers and exclusions}
\label{sec:modifiers}

Two other optional, mutually exclusive modifiers further couple the otherwise constant rates to the state of the system.
The first enables us to incorporate the cooperative melting of a binder such as DNA~\cite{mitra2023coarse}. 
A sigmoidal switch
\begin{equation}
\label{eq:gT}
g(T) = \tfrac{1}{2}\left[\tanh\!\left(\lambda (T - T_\mathrm{melt})\right) + 1\right]
\end{equation}
interpolates each rate between a low-temperature and a melted value,
\begin{equation}
\label{eq:kT}
k_{\mathrm{on}/\mathrm{off}}(T) = k^{\mathrm{init}}_{\mathrm{on}/\mathrm{off}}\left[1 - g(T)\right]
+ k^{\mathrm{melt}}_{\mathrm{on}/\mathrm{off}}\,g(T),
\end{equation}
with $\lambda$ setting the sharpness of the transition. Choosing
$k^\mathrm{melt}_\mathrm{on} = 0$ forbids binding above $T_\mathrm{melt}$, and
$k^\mathrm{melt}_\mathrm{off}$ then fixes the bound fraction at the transition;
requiring half of the pairs to be bound at $T_\mathrm{melt}$ determines this constant
uniquely~\cite{mitra2023coarse}.

We have also included, as an example, a load-dependent acceleration of unbinding (a slip bond) in the manner of Bell's model for adhesion molecules~\cite{bell1978models},
\begin{equation}
\label{eq:slip}
P_\mathrm{off}(r) = P^{0}_\mathrm{off}\,
\exp\!\left[a\sqrt{\frac{U_\mathrm{b}(r)}{k_\mathrm{B}T}}\right].
\end{equation}
The square root is what makes this a force criterion rather than an energy one:
for a harmonic bond the tension is $f = \sqrt{2k_\mathrm{b}U_\mathrm{b}}$, so the
exponent is $f x^{*}/k_\mathrm{B}T$ with a Bell distance
$x^{*} = a\sqrt{k_\mathrm{B}T/2k_\mathrm{b}}$ proportional to the thermal
fluctuation $\sqrt{k_\mathrm{B}T/k_\mathrm{b}}$ of the bond length. A single
parameter $a$ thus tunes how strongly a stretched bond is destabilized, which is
the ingredient needed to model slip bonds under load in a rheological
measurement.

DyBond also supports intramolecular exclusions; sites belonging to the same molecule---identified
once at startup by tracing the static, unchanging bond topology---can be excluded from reacting with each other, so that a building block cannot bond to itself. The exclusion has no default: each reaction must state whether it applies, through \texttt{selfAvoidFlag}, both in the DyBond plugin (\cref{fig:dybond_api}) and in pySNAP's \texttt{params/dybond.yaml} (\cref{fig:dybond}). Disabling the exclusion admits intramolecular bonding, allowing self-loops---one of the topological defects that limit the elasticity of real networks---to form between the sites of a single molecular building block.

\subsection{GPU implementation}
\label{sec:gpu}

The sweep of \cref{sec:sweep} runs either on the host or entirely on the accelerator, with the device path written against the Heterogeneous-computing Interface for Portability (HIP) so that it compiles for both CUDA and ROCm. The four phases map to four GPU kernels launched in series, which functionally act as barriers so that occupancies are read-only while candidates are being searched and are changed only in the matching kernel. Therefore, a pair is claimed by two atomic compare-and-swap operations on the shared occupancy array, always attempted on the lower particle tag first and rolled back if the second fails, which prevents both double occupancy and half-claimed sites without a global lock. Only the compacted lists of formed and broken bonds return to the host, where they are applied to HOOMD-blue's bond table.

Mirroring HOOMD-blue's architecture, randomness is drawn from a counter-based Philox generator~\cite{salmon2011parallel}, keyed by the simulation seed, the current time step, and an identifier distinguishing the binding and unbinding streams of each reaction channel, and counted by the sorted pair of particle tags. The random number governing a given trial is therefore a pure function of when and which pair is evaluated, independent of thread scheduling, particle memory layout, or domain decomposition. Consequently, every breaking decision and every acceptance evaluation for a given candidate pair is identical across the CPU and GPU implementations.

The only operational divergence occurs during the matching phase under partner contention. The CPU considers the candidate sites of a reaction in a pseudo-random permutation redrawn every sweep, whereas the GPU resolves competing claims through hardware-level atomic compare-and-swap (\texttt{atomicCAS}) operations over a fixed ordering of the candidate list, so the claim that prevails is set by the order in which those operations complete. That order is arbitrary but not randomized, and unlike the CPU permutation it is not guaranteed to treat competing sites evenhandedly. The two paths therefore agree exactly for every uncontested trial and can differ statistically only when several free sites compete for the same partner. The molecule-exclusion tests of \cref{sec:validation} run once per available device, so a divergence between the two paths in that bookkeeping fails the test suite on any host that has both.

Because the bonding kernels consume the neighbor list the integrator has already
built, the additional cost of dynamic bonding is one pass over the neighbors of
the reactive sites every $n$ steps. Reactive sites are typically a small
fraction of the particles and $n$ is typically large, so the sweep is a minor
addition to the cost of the underlying MD.

\subsection{DyBond Setup and HOOMD Integration}

\begin{figure}[t]
\includegraphics{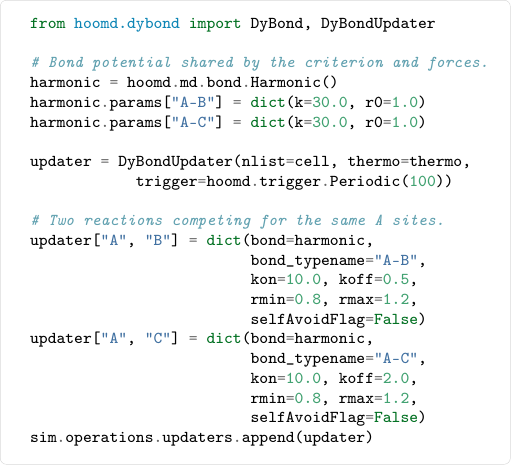}
\caption{Example of registering two competing reactions with the DyBond plugin. Sites of
type \texttt{A} may bond to either \texttt{B} or \texttt{C}, but hold at most
one dynamic bond at a time, so the two reactions compete through the shared
occupancy record of \cref{sec:sweep}. Here the fourfold difference in
$k_\mathrm{off}$ makes the \texttt{A}--\texttt{B} product the more stable of the
two by $k_\mathrm{B}T \ln 4$. Each reaction takes its rest length $r_0$ from
the bond potential it names and must state, through \texttt{selfAvoidFlag},
whether sites of the same molecule may bond to each other. The neighbor list
cutoff for each reacting type pair is raised to $r_\mathrm{max}$, if it is
smaller, when the updater attaches, and the trigger period supplies the $n$ of
\cref{eq:probs}.}
\label{fig:dybond_api}
\end{figure}

The user interface of \texttt{DyBond} intentionally mirrors HOOMD-blue's modular simulation setup methodology, adopting its type-pair dictionary syntax, trigger-governed scheduling, and composable operations. As shown in \cref{fig:dybond_api}, the plugin is implemented as a standard \texttt{hoomd.operation.Updater} appended to \texttt{sim.operations.updaters} to execute periodically alongside the numerical integrator. To avoid redundant parameter definitions, the underlying bonded potential (e.g., \texttt{Harmonic}) is instantiated once and shared between the MD force compute and the \texttt{DyBondUpdater}, ensuring the acceptance criterion evaluates the exact potential energy governing particle dynamics.

Following HOOMD's parameter assignment conventions, individual reactions are registered on the updater using tuple keys of particle types (e.g., \texttt{updater[``A'', ``B'']}). The updater automatically manages downstream dependencies: it queries the attached \texttt{ThermodynamicQuantities} compute for the temperature needed in the Boltzmann factor and updates the neighbor list cutoffs, $r^\alpha_{\mathrm{cut}}$, to $r^\alpha_{\mathrm{max}}$ for each reacting pair if $r^\alpha_{\mathrm{cut}} < r^\alpha_{\mathrm{max}}$. The trigger period directly supplies the step interval $n$ in \cref{eq:probs}, so that each bonding sweep covers a time $n\,\Delta t$ without requiring manual rate rescaling.

Crucially, multiple reactions can target overlapping particle types without additional user configuration. In \cref{fig:dybond_api}, sites of type \texttt{A} can form dynamic bonds with either \texttt{B} or \texttt{C}. Because all registered reaction channels share the particle occupancy record described in \cref{sec:sweep}, monovalency is strictly enforced: an \texttt{A} site claimed by one reaction is masked from the other. Differences in kinetic parameters—here a fourfold lower $k_\mathrm{off}$ for \texttt{A}--\texttt{B} compared to \texttt{A}--\texttt{C}—naturally translate into the expected $k_\mathrm{B}T \ln 4$ free energy difference between the competing products.

\section{pySNAP: Fundamentals and features}
\label{sec:use}
pySNAP is available from \url{https://github.com/hocky-research-group/pySNAP} with documentation provided at \url{https://hockygroup.com/pySNAP/}, and is distributed as two primary packages: \texttt{snap\_workflow} and \texttt{snap\_simulate}. \texttt{snap\_workflow} is a general-purpose simulation pipeline manager and command-line tool with no dependency on either HOOMD-blue or any other pySNAP packages, so it can be adopted and installed independently; \texttt{snap\_simulate} is a simulation setup tool that pairs with \texttt{snap\_workflow} to turn parameter files into initial configurations and configured HOOMD \texttt{Simulation} objects, while also providing driver routines for density compression and autocorrelation minimization.

\subsection{Declarative Workflow Primitives and SPoints}
\label{sec:workflow_primitives}

\begin{figure}[h]
    \centering
    \includegraphics[width=\columnwidth]{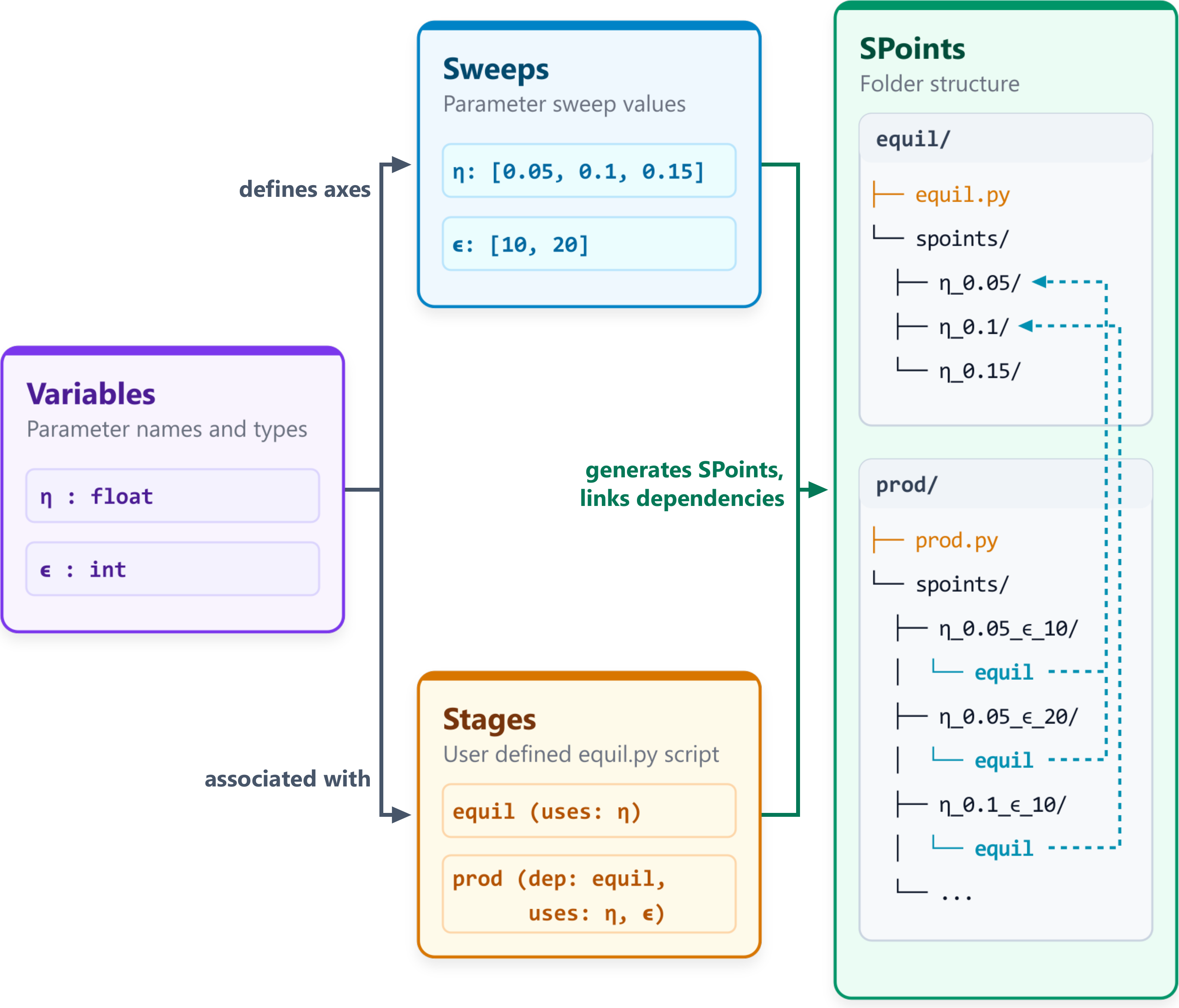}
    \caption{Architecture of pySNAP’s declarative workflow engine. Pipelines are defined by typed Variables and discrete parameter Sweeps, whose Cartesian product produces the parametric space. Simulation Stages are user-defined tasks topologically sorted by execution dependencies. The engine maps parameter combinations into discrete SPoints (state points) and automatically links upstream stage artifacts to downstream stages for unified access across SPoints.}
    \label{fig:workflow_relationship}
\end{figure}

Systematic exploration of dynamic bonding phenomena requires surveying a combinatorial parameter space—spanning association rates ($k_\mathrm{on}$), bond strengths ($\varepsilon$), and patch geometries. Managing multi-stage simulations across hundreds of state points with ad hoc shell scripts is error-prone, hard to audit, and hinders computational reproducibility. To eliminate manual intervention, pySNAP formalizes computational sweeps around three declarative primitives—\textit{Variables}, \textit{Stages}, and \textit{Sweeps}—from which the execution engine deterministically derives and materializes discrete execution targets termed \textit{SPoints} (\cref{fig:workflow_relationship}). To streamline pipeline setup, all workflow primitives can be declared through an interactive command-line wizard (\texttt{workflow <primitive> add}).

The parameter space is formalized by declaring typed \textit{Variables} (such as integer bead counts or floating-point reaction rates). Enforcing explicit data types ensures strict schema validation across the entire workflow, preventing formatting ambiguities or floating-point discrepancies from silently corrupting configuration files or filesystem paths.

The simulation workflow is decomposed into modular computational phases termed \textit{Stages}. In a typical MD study, stages correspond to sequential operations such as initial mixture assembly, density compression, thermal equilibration, and production sampling. Each stage encapsulates its own simulation logic and explicitly declares the subset of variables it requires at creation. Decoupling stages by variable requirements ensures that an upstream phase (such as density compression) is not unnecessarily re-run across parameter variations meant exclusively for downstream production sampling.

Experimental design is enabled through systematic parameter \textit{Sweeps} configured by assigning discrete value sequences for each \textit{Variable}. The Cartesian product of active \textit{Sweep} axes at each \textit{Stage} defines the set of parameter combinations to be explored. Each individual parameter combination at each stage defines an individual \textit{SPoint}.  

pySNAP automatically materializes each conceptual SPoint as a deterministic directory within the filesystem hierarchy (e.g., \texttt{prod/spoints/$\eta$\_0.05\_$\varepsilon$\_10/}). Each SPoint directory contains and isolates all its own logs and user-defined simulation outputs. Deterministic, coordinate-based naming ensures that all simulation artifacts remain accessible across sweeps and prevents redundant simulation runs when distinct sweeps contain overlapping parameter space.

Inter-stage dependencies are managed by parsing stage relationships into a Directed Acyclic Graph (DAG). By evaluating the DAG through a topological sort, pySNAP establishes strict execution order—preventing dependent child \textit{SPoints} from running prior to parent completion—while automatically resolving inter-stage data dependencies by generating symbolic links to corresponding parent SPoints (illustrated by the dashed links in \cref{fig:workflow_relationship}). Consequently, downstream simulation scripts can access predecessor artifacts—such as equilibrated restart configurations—through predictable local paths, eliminating hard-coded filesystem references and error-prone runtime path resolution.

\subsection{Stage Implementation, Parameters, and Execution Context}
\label{sec:stage_and_env}

To avoid the monolithic entanglement of parameter values, runtime environments, and simulation logic that so often plagues computational workflows, pySNAP enforces a strict separation within each stage. Rather than embedding physical constants, cluster submission directives, or argument parsing boilerplate directly into simulation code, each stage isolates concerns into three distinct components: an externalized parameter specification (\texttt{params/}), a pure simulation task script (\texttt{<stage>.py}), and an independent execution environment (\texttt{env.yaml}). This  modularity ensures that researchers can immediately identify and adjust the relevant component when expanding a workflow to new use cases—whether exploring new physical parameters, modifying simulation protocols, or transferring the workflow to new compute architectures—without having to parse mixed-concern scripts.

\begin{figure}[t]
\includegraphics{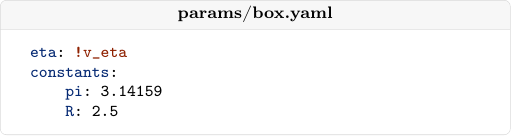}
\caption{Example parameter declarations within the \texttt{params/} directory. The specified SPoint values of each variable are injected in-place at variable directives (denoted by \texttt{!v\_variable}) before the SPoint is executed. Nested dictionaries and lists are allowed.}
\label{fig:params_decl}
\end{figure}

\begin{figure}[!htbp]
\includegraphics{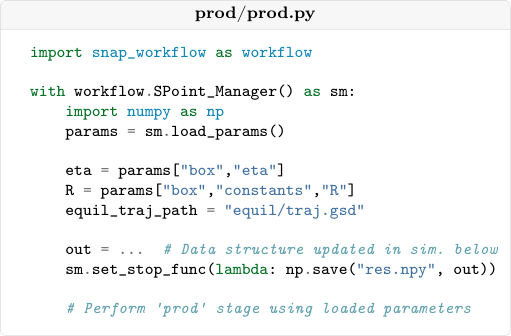}
\caption{Example stage simulation script demonstrating \texttt{SPoint\_Manager} use, simulation checkpointing, parameter and variable fetching, and dependency access. The stage simulation script is run from within each specific SPoint directory, allowing direct access to dependency links created during SPoint generation.}
\label{fig:stage_script}
\end{figure}

Physical constants, force-field parameters, and algorithm settings are declared in human-readable YAML (or JSON) configurations under the \texttt{params/} directory, where files and subdirectories automatically map into a unified, nested parameter tree. Rather than requiring bespoke argument parsers to inject sweep values, pySNAP uses variable directives (\texttt{!v\_variable}, \cref{fig:params_decl}) that can be placed at arbitrary depths within these nested dictionaries and lists. During SPoint initialization, the engine performs in-place substitution of active parameter coordinates into matching directives. The resulting structure is then easily accessed in simulation scripts and parsed using clean multidimensional key indexing (\cref{fig:stage_script}), eliminating manual file loading and command-line argument parsing.

The computational work of each stage is defined in a standalone Python script (e.g., \texttt{prod.py}, \cref{fig:stage_script}). Lifecycle tracking is managed by the \texttt{SPoint\_Manager} context manager, which handles execution state monitoring (e.g., initialized, running, complete, error), exposes the \texttt{load\_params()} method for type-safe parameter retrieval, and provides \texttt{set\_stop\_func()} for automated shutdown checkpointing. pySNAP executes the script directly inside each isolated SPoint working directory, allowing read and write operations to target simple relative paths—including accessing dependent SPoint data (e.g., \texttt{equil/traj.gsd}) via the symbolic links created during SPoint generation. This design eliminates runtime path construction and decouples stage logic from global filesystem layouts.

\begin{figure}[!htbp]
\includegraphics{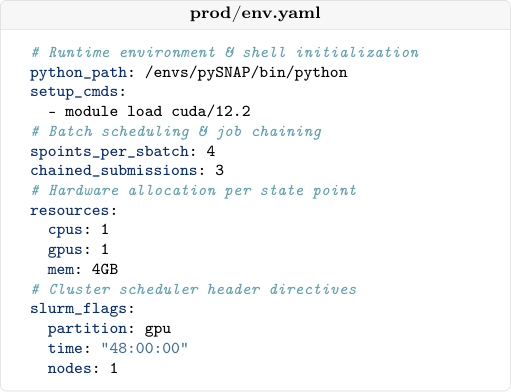}
\caption{Stage runtime environment and execution specification. Each pipeline
stage isolates its execution context through an \texttt{env.yaml} configuration.
The \texttt{python\_path} and \texttt{setup\_cmds} guarantee reproducibility
by loading necessary system modules and targeting a dedicated virtual
environment. \texttt{resources} defines the per-state-point hardware footprint
used for dynamic process packing on local workstations. For cluster deployment,
\texttt{spoints\_per\_sbatch} aggregates multiple parameter runs into single
batch allocations to avoid queue throttling, while \texttt{chained\_submissions}
automatically submits sequential continuations using Slurm job dependencies for long simulations that exceed wall time limits.}
\label{fig:stage_env}
\end{figure}

Finally, execution requirements and hardware configurations are decoupled from simulation scripts via \texttt{env.yaml} (\cref{fig:stage_env}). Computational reproducibility across heterogeneous compute targets is guaranteed by specifying the explicit Python interpreter (\texttt{python\_path}) and shell environment hooks (\texttt{setup\_cmds}), ensuring necessary system modules (such as CUDA) are loaded and environment variables are exported prior to execution. Although pySNAP also supports dynamic resource dispatching through alternate execution modes—using the \texttt{resources} block to pack concurrent jobs onto local hardware or task-independent node reservations—these capabilities are not discussed in detail here.

To account for individual cluster Slurm configurations, pySNAP takes Slurm \texttt{\#SBATCH} directives directly under \texttt{slurm\_flags} and passes them as command-line arguments during submission. To minimize scheduler overhead and avoid queue throttling, \texttt{spoints\_per\_sbatch} aggregates multiple independent SPoints into a single batch job allocation. Additionally, because long-timescale dynamic bonding simulations often exceed standard cluster wall time limits, \texttt{chained\_submissions} automatically submits sequential continuation jobs linked by Slurm job dependencies to ensure multi-day simulation trajectories advance continuously from restart checkpoints without requiring manual monitoring or resubmission.

\subsection{Automated Simulation Compilation}
\label{sec:sim_compilation}

In standard MD workflows, researchers frequently write procedural Python scripts spanning hundreds of lines to configure devices, neighbor lists, pair potentials, rigid bodies, and file writers. In pySNAP, this boilerplate is replaced by an automated simulation configuration tool (\texttt{snap\_simulate}) that translates declarative parameter files directly into HOOMD-blue \texttt{Simulation} objects.

\begin{figure}[!htbp]
\includegraphics{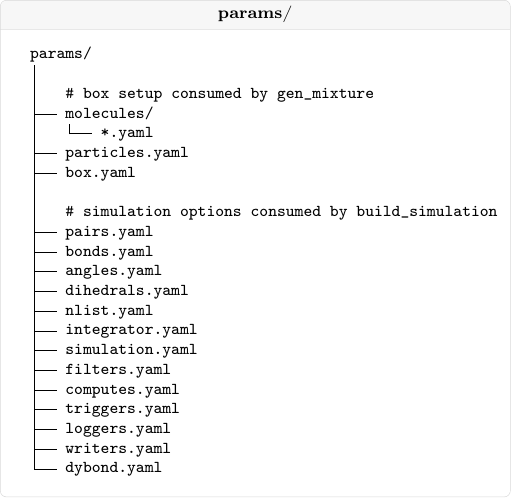}
\caption{Modular parameter schema serving as an \textit{a la carte} catalog of available simulation control points. System generation routines (\texttt{gen\_mixture}) consume particle, molecular, and box specifications to assemble initial \texttt{.gsd} coordinate frames. The simulation compiler (\texttt{build\_simulation}) parses the remaining interaction, integration, and diagnostic configurations to instantiate an executable HOOMD-blue \texttt{Simulation} object. Dynamic bonding is enabled by the presence of \texttt{dybond.yaml}.}
\label{fig:params}
\end{figure}

The complete physical specification is organized within the \texttt{params/} directory (\cref{fig:params}). Parameter files fall broadly into two functional groups: structural specifications consumed during initial state generation (such as \texttt{box.yaml}, \texttt{molecules/}, and \texttt{particles.yaml}, processed by \texttt{snap\_simulate.system.gen\_mixture}), and runtime physics specifications consumed during simulation assembly (such as \texttt{integrator.yaml}, \texttt{pairs.yaml}, and \texttt{dybond.yaml}, processed by \texttt{snap\_simulate.hoomd.build\_simulation}). Crucially, the default behavior for any omitted file is simply that the corresponding HOOMD object is not instantiated. In fact, the only strict requirements for \texttt{build\_simulation} are the compute device and seed specifications (\texttt{simulation.yaml}), the numerical integration method (\texttt{integrator.yaml}), and the particle type declarations (\texttt{particles.yaml})—with all interaction potentials, diagnostic computes, and updaters acting as optional attachments. The tree in \cref{fig:params} thus functions as an \textit{a la carte} menu of modular control points rather than an exhaustive set of mandatory inputs.

To assemble initial coordinates, \texttt{gen\_mixture} parses the structural specifications—namely \texttt{particles.yaml}, \texttt{box.yaml}, and parameterized molecular schemas in \texttt{molecules/}—into a valid starting configuration. Target box dimensions and densities are computed automatically from specified packing or volume fractions. Constituent species can be procedurally generated from a library of built-in parametric templates (including linear, ring, and branched polymers, as well as patchy or grafted colloids) or explicitly specified as custom molecular architectures with user-defined bead coordinates and local connectivity. To prevent unphysical particle overlaps during dense packing, \texttt{gen\_mixture} utilizes a cell-list spatial partition approach that validates placements across random or lattice-based insertion schemes. Crucially, the routine automates both topological assembly and rigid-body initialization: for bonded architectures, it indexes global bond, angle, and dihedral arrays; for rigid bodies, it calculates principal moments of inertia, assigns quaternion orientations, and reorders constituent particles to satisfy HOOMD-blue's internal conventions. The assembled configuration can be returned directly as an in-memory \texttt{gsd.hoomd.Frame} object for immediate integration or serialized to an initial \texttt{.gsd} snapshot file, cleanly decoupling initial state generation from runtime physics compilation.

Individual components are declared in modular YAML files and wired together through a category-scoped name-referencing system. When configuring dependent components, supporting objects—such as neighbor lists, computes, filters, or triggers—are specified simply by their user-defined string keys rather than instantiated Python objects. For example, a pair potential specifies a neighbor list defined in \texttt{nlist.yaml} by name (e.g., \texttt{nlist: cell}), while a trajectory writer references a periodic cadence defined in \texttt{triggers.yaml} (e.g., \texttt{trigger: periodic\_100}). At build time, \texttt{snap\_simulate} automatically resolves each key within its corresponding object registry, allowing complex simulation networks to be assembled without procedural Python plumbing.

\begin{figure}[t]
\includegraphics{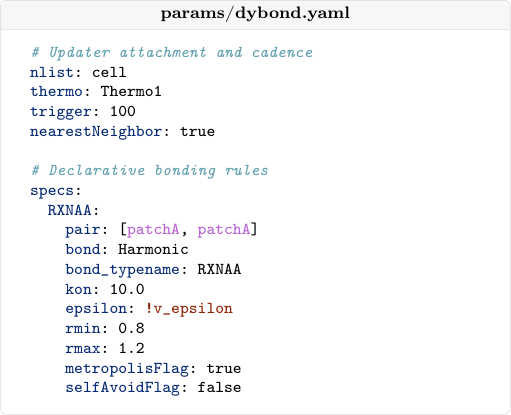}
\caption{Specification of the dynamic bonding updater (\texttt{params/dybond.yaml}). Two particles of type \texttt{patchA} within the distance window $[r_\mathrm{min}, r_\mathrm{max}]$ can form a \texttt{Harmonic} bond of type \texttt{RXNAA}, subject to a Metropolis acceptance test evaluated at the period set by \texttt{trigger}. The sweep variable \texttt{epsilon} is substituted in-place via \texttt{!v\_epsilon}. Additional reaction rules under \texttt{specs} enable multi-component bonding networks.}
\label{fig:dybond}
\end{figure}

In particular, this architecture natively automates the setup of \texttt{DyBond} (\cref{fig:dybond}). Rather than requiring researchers to manually instantiate potentials, configure updaters, and register reaction channels in Python as shown in \cref{fig:dybond_api}, \texttt{snap\_simulate} configures a complete \texttt{DyBondUpdater} directly from \texttt{params/dybond.yaml}. The configuration header binds the updater to the active neighbor list (\texttt{nlist: cell}) and thermodynamic compute (\texttt{thermo: Thermo1}) declared in their respective files in \texttt{params/}, while the \texttt{specs} block maps directly to individual reaction channels. Crucially, declaring these channels in YAML integrates them seamlessly with the workflow engine: the bond energy is assigned via a sweep directive (\texttt{epsilon: !v\_epsilon}), enabling automated exploration of bonding thermodynamics across SPoints without altering the underlying simulation script. Furthermore, \texttt{snap\_simulate} automatically enforces physical closure relations during compilation: specifying any two of $\{k_\mathrm{on}, k_\mathrm{off}, \varepsilon, P^{0}_\mathrm{on}\}$ is sufficient for the compiler to deduce the rest through \cref{eq:epsilon,eq:probs}, except the pair $k_\mathrm{on}$ and $P^{0}_\mathrm{on}$, which \cref{eq:probs} makes equivalent; an over-specified set is checked for consistency.

Inside a stage script, calling \texttt{sim, hoomd\_objects = build\_simulation(params)} compiles the requested components and returns both the active \texttt{hoomd.Simulation} instance and a dictionary containing references to all created objects (e.g., neighbor lists, integrators, filters, and updaters). Rather than functioning as an inflexible black box, this design allows researchers to dynamically modify parameters in memory prior to compilation or attach bespoke forces, hooks, or custom runtime callbacks after compilation. For instance, a multi-phase stage script can easily perform non-bonded thermal equilibration prior to initiating reaction dynamics simply by popping \texttt{dybond} from \texttt{params} before building the initial simulation object.

\subsection{Workflow Execution and State Monitoring}
\label{sec:submit_and_status}

Job dispatching is driven from the command line using \texttt{workflow spoint submit}. Rather than blindly submitting all declared SPoints, the submission engine queries the stage DAG and evaluates the status of every individual SPoint. An SPoint is dispatched to the Slurm scheduler only if it has an available status (i.e., \texttt{initialized} or \texttt{incomplete}) and all of its parent stage prerequisites are marked \texttt{complete}. The submission engine attempts to submit as many SPoints as are ready and respects \texttt{spoints\_per\_sbatch}, aggregating multiple independent SPoints from the same stage into a single batch allocation to maximize resource utilization and prevent scheduler queue throttling.

During execution, the real-time progress of the entire campaign is monitored using \texttt{workflow status}. Each SPoint transitions through well-defined lifecycle states: \texttt{initialized}, \texttt{submitted}, \texttt{running}, \texttt{complete}, \texttt{incomplete}, or \texttt{error}. Because downstream stages are strictly gated by parent completion, dependent runs automatically become eligible for submission as soon as their prerequisite SPoints finish and flush checkpoints to disk. If a node failure or stage script results in an error, the affected SPoint can be diagnosed and resubmitted in isolation without re-running completed branches of the DAG, providing robust fault isolation across large-scale parametric campaigns. SPoints that time out before completion are marked as \texttt{incomplete} and will be submitted again during the next \texttt{workflow spoint submit} command invocation. Note that \texttt{incomplete} SPoints will likely not complete on subsequent submissions unless the stage script is configured to load from and save to checkpoints.

\section{Usage Examples}
\label{sec:examples}

This section works through three examples of increasing scope. The complete
input files for all three, including parameter directories and sweep
definitions, are available in a companion repository,
\url{https://github.com/hocky-research-group/pySNAP-paper} (see Data Availability).
All three are simulated in reduced units: lengths in units of the bead diameter $\sigma$,
masses in units of the bead mass $m$, energies in units of $k_\mathrm{B}T$, and times in
units of $\tau = \sigma\sqrt{m/k_\mathrm{B}T}$. The integration time step is
$\Delta t = 0.001\tau$ in the first two examples and $0.0005\tau$ in the third, and rate
constants such as $k_\mathrm{on}$ are given in units of $\tau^{-1}$.
Each example is equilibrated with dynamic bonding off and then run in production:
$5\times10^{5}$ and $8\times10^{7}$ steps ($500\tau$ and $8\times10^{4}\tau$) in the first two
examples, and $2\times10^{5}$ and $4\times10^{7}$ steps ($100\tau$ and $2\times10^{4}\tau$) for the
droplets of \cref{fig:droplet}.

\subsection{A self-assembling colloid}
\label{sec:ex_colloid}

\begin{figure}[!htbp]
\includegraphics{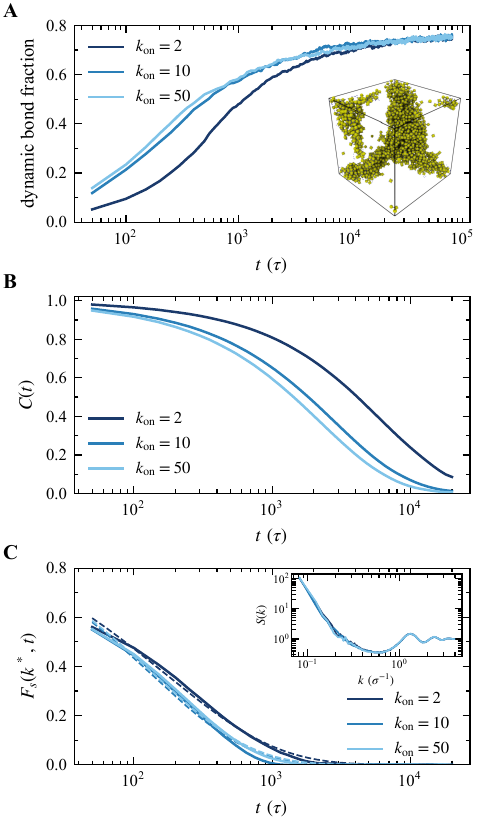}
\caption{Patchy colloid assembly at different bonding rates with fixed $\varepsilon=8$. (A) Fraction of dynamic bonds formed versus simulation time. Inset: snapshot of an equilibrium gel formed with $k_\mathrm{on}=50$. (B) Bond correlation function versus time interval. (C) Self-intermediate scattering functions of the colloidal center particles at $k^*=1.334\sigma^{-1}$. The dashed lines are stretched exponential fits to the data. The fitted exponents are 0.54, 0.62 and 0.61 with increasing $k_{\rm on}$. Inset: static structure factor calculated from 60 frames in the time range [60000, 80000]$\tau$. The bonding is triggered every 100 steps for $k_\mathrm{on}=2, 10$  and every 10 steps for $k_\mathrm{on}=50$ so that $k_\mathrm{on}\,n\,\Delta t \le 1$ in every run.}
\label{fig:NC}
\end{figure}

We begin with perhaps the simplest building blocks containing specific, reversible bonding, a one-component patchy
colloid whose attractive surface sites bond reversibly to one another. This example shows the
whole path from parameter files to assembled gels---system generation, the
two-stage equilibration and production sweep, and the analysis of resulting trajectories on structures and dynamics---on a system
small enough to follow completely.

We simulate the system with 2000 colloidal particles, each with six patches for dynamic bonding. The patches are constrained to the surface by FENE \cite{warner1972kinetic} bonds plus a WCA \cite{weeks1971role} potential and are constrained to an octahedral arrangement by a strong cosine-squared angular potential. We fix the volume fraction $\eta=0.05$ and the dynamic bond energy $\varepsilon=8$ in units of $k_{\rm B}T$. At this constant bond strength, we sweep the bonding rate constant (``kon'' in \cref{fig:dybond}).

The simulations start with a short first pre-equilibration stage without turning on dynamic bonds, followed by the production stage. The bonding rate controls kinetics, while all gels converge to approximately the same equilibrium state by the end of the simulations (\cref{fig:NC}A).
As the bonding rate increases, gelation crosses over from reaction-limited to diffusion-limited kinetics, beyond which the gelation rate saturates: raising $k_\mathrm{on}$ fivefold from 2 to 10 more than doubles the bond fraction formed in the first $50\tau$, while a further fivefold increase to 50 raises it by less than a fifth. The bond correlation function quantifies the fraction of dynamic bonds present at a reference time that remain intact after a time interval $t$ (\cref{fig:NC}B). 
Since $k_{\rm off}$ scales with $k_{\rm on}$ at fixed bond energy, increasing $k_{\rm on}$ also accelerates bond turnover, leading to a faster decay of bond correlation.
The systems form phase-separated gels characterized by a high peak in the low-$k$ regime of the static structure factor of the colloidal particle centers (\cref{fig:NC}C), regardless of the bonding rate. We further calculate the self-intermediate scattering function $F_s(k^*,t)$ at the primary peak $k^*$ of $S(k)$ in \cref{fig:NC}C. 
The curves decay to zero within a short timescale, much shorter than the bond persistence time, and are insensitive to the bonding rate. This indicates that the main motion in the systems is gel fluctuation without breaking dynamic bonds. The fitted exponents from the stretched exponential functions range from 0.54 to 0.62, indicating sub-diffusive particle motion.

\subsection{A mixed tetraPEG and colloid network}
\label{sec:ex_mixed}

\begin{figure*}[ht]
\includegraphics{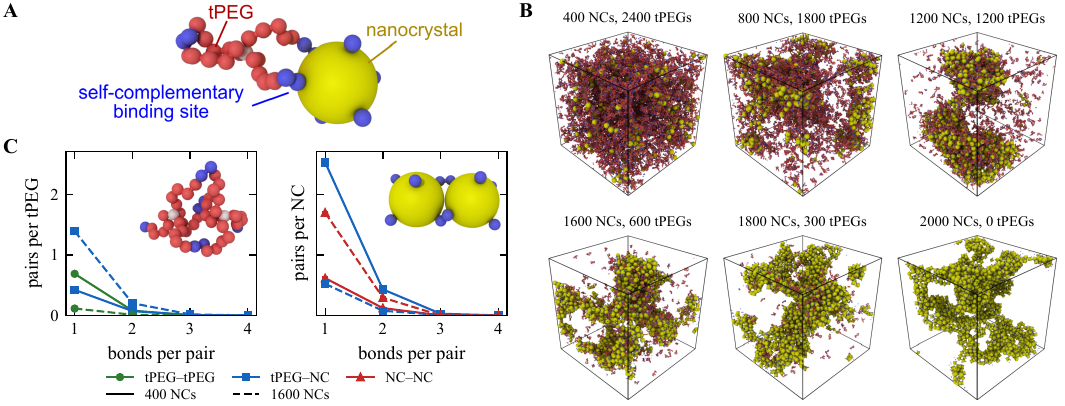}
\caption{The mixtures of tetraPEG and nanocrystals. (A) Self-complementary patches can form both intramolecular bonds and intermolecular bonds. (B) Last-frame snapshots of the simulation boxes across the composition sweep, with NCs in yellow, tPEG beads in red, and binding sites in blue. The gel morphology transitions from relatively homogeneous to increasingly inhomogeneous as the NC fraction increases. (C) Bond multiplicity for different molecular pairs: the number of pairs joined by one to four dynamic bonds, per tPEG (left) and per NC (right), for systems with 400 (solid) and 1600 (dashed) NCs. Insets: a triple-bonded tPEG pair (left) and a double-bonded NC pair (right).}
\label{fig:composite}
\end{figure*}

The second example is a composite network combining patchy
colloids with four-armed star polymers that compete for the same reactive
sites. The point of the example is how little separates it from the first in terms of the pySNAP setup.
The building blocks are two separate entries in the molecules directory, and the composition
becomes one more sweep variable. However, neither the driver script nor the bonding specification needs any changes. 

We consider mixtures of tetraPEGs (tPEGs) and nanocrystals (NCs), where the NCs are modeled identically to the colloids in the first example.  Each tPEG molecule consists of four flexible six-bead arms, including terminal patches that can form dynamic bonds with NCs, other tPEGs, or other patches within the same molecule (\cref{fig:composite}A). We study a series of systems with a fixed total number of patches and the total volume fraction $\eta=0.05$. We observe increasing inhomogeneity in the gel morphology with increasing NC fraction in \cref{fig:composite}B. The bond statistics and network connectivity also vary with the molecular ratio, with a noticeable fraction of molecular pairs connected by multiple dynamic bonds (\cref{fig:composite}C).

\subsection{DNA-coated emulsion droplets}
\label{sec:ex_droplets}

\begin{figure*}[ht]
\includegraphics[]{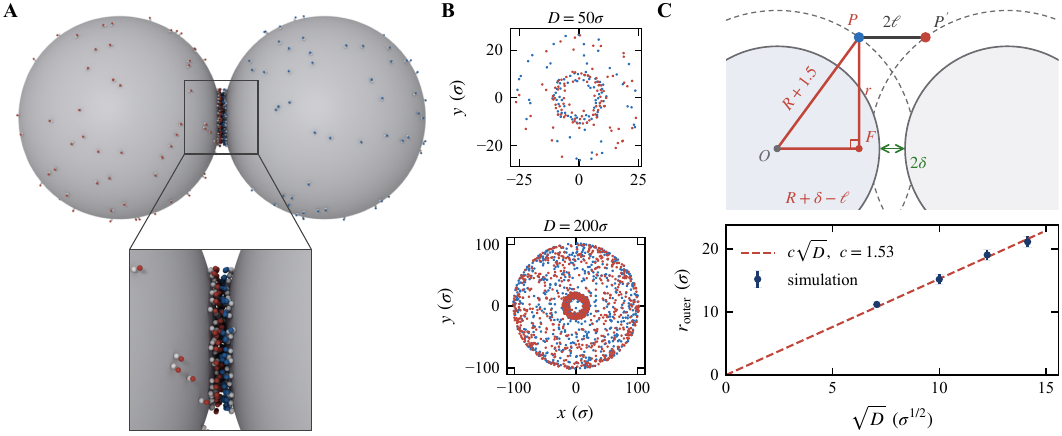}
\caption{DNA-coated emulsion droplets form ring-shaped patches. (A) A snapshot of a pair of droplets with diameter $D = 100\sigma$, each with 200 binding sites (in blue or red) and 200 anchor particles (in white). Both the binding sites and anchor particles have diameter $\sigma$. Dynamic bonds form between the blue and red complementary binding sites. Inset: side view of the boxed contact region, where the bonded binding sites form the adhesion ring. (B) Binding sites projected onto the plane orthogonal to the droplet center-to-center axis, in the last frame of production, for droplets with $D = 50\sigma$ and 100 binding sites each (upper) and with $D = 200\sigma$ and 800 binding sites each (lower). The dense inner ring is the adhesion patch; the sites scattered around it lie elsewhere on the droplet surfaces. (C) Upper: Geometric relationships between the outer radius $r$ and the droplet radius $R$. OP is the distance from the droplet center to the binding site. $2l$ is the equilibrium length of dynamic bonds, set as $1.97\sigma$ in the simulations. $2\delta$ is the surface-to-surface separation between droplets. Lower: The outer radius exhibits a linear relationship to the square root of the droplet diameter $D$; the dashed line is the fit $r = c\sqrt{D}$ with $c = 1.53$. In each frame the outer radius is the mean distance from the center-to-center axis of the five outermost bonded binding sites; it is averaged over the final 50 saved frames, and the error bars are the standard deviation over those frames.}
\label{fig:droplet}
\end{figure*}

The third example deliberately leaves the linker-mediated gels of
\cref{sec:ex_colloid} and \cref{sec:ex_mixed} behind, to show that the framework can be easily adapted to other self-assembling soft-matter systems of interest. To do so, we re-implement a model for colloidal droplets with explicit mobile DNA binders \cite{mcmullen2018freely} studied in Ref.~\cite{mitra2023coarse}.
In this system, binders diffuse on a droplet surface and bond to complementary binders on a neighboring droplet, concentrating into
adhesion patches whose size sets the droplet valence. The geometry, the length  scale, and the physical question all differ from the preceding examples, but the description is again a directory of parameter files, and the bonding is again one simple reaction specification.

We simulate a pair of emulsion droplets bearing complementary binding sites that are tethered to the droplet surfaces through anchor particles that are themselves bonded to the center of the droplet, allowing free motion. Harmonic angular potentials cause the binding sites to stand vertically from the surface. To speed up the simulations, the anchors of each droplet are bonded to a rigid core of coincident beads at its center rather than to a single central particle, which leaves the forces unchanged (\cref{sec:rigid_core}). The binding sites and anchors are initially distributed according to spherical Fibonacci patterns but subsequently redistribute as bonds form, producing a ring-shaped adhesion patch between the droplets (\cref{fig:droplet}A) at both the smallest and the largest diameters studied (\cref{fig:droplet}B). Although individual binding sites within the patch continuously fluctuate, the outer radius of the patch remains relatively stable because of the geometric constraint imposed by the two dynamically bonded droplets (\cref{fig:droplet}C). Simulations over four droplet diameters reveal that the patch outer radius scales proportionally to the square root of the droplet diameter, consistent with the prediction from geometry in~\cref{sec:geometry}. This geometry is experimentally relevant: the large disparity between the sizes of the droplets and binding sites is reflective of the disparity between micrometer-scale emulsion droplets and the DNA anchored to the surface. These ring-shaped distributions of binding sites have been observed experimentally at contacts between DNA-coated droplets and substrates~\cite{judd2025statistical}.
\section{Conclusions}
\label{sec:conclusions}

We have described pySNAP, a set of tools for simulating soft materials whose
building blocks assemble through reversible bonding.
The workflow layer takes the description of a
simulation out of a script and into a directory of parameter files,
so that a multi-dimensional sweep is constructed by the workflow engine rather than by hand through bespoke scripts.
The resulting state points are staged, submitted, packed onto available resources, and collected by one
command-line tool. 
The bonding layer, DyBond, generalizes the dynamic bonding
method of Ref.~\cite{mitra2023coarse} in three directions: an arbitrary number
of chemically distinct reactions may compete for a shared pool of reactive
sites, the Metropolis acceptance factor is evaluated with whichever bond
potential the integrator uses rather than an assumed harmonic form, and the
whole bonding sweep can run on the GPU with stochastic decisions that are
independent of thread scheduling. The examples of \cref{sec:examples} carry
these through a single-component patchy colloid, a two-component polymer and
nanocrystal network, and a reproduction of a published droplet model built on
entirely different geometry.

As implemented, there remain several limitations to the current approach.
First, we only allow a reactive site to carry at most one dynamic bond at a time. 
This is the right model for our level of coarse-graining where we use a single bead to represent the reactive chemical group, but means that multiple bonding between building blocks must be implemented through explicit sites.

The scheme is also purely dissociative: a bond must break before either partner
can bond again, and the reaction that broke it cannot re-form it within the same
sweep. We have not implemented associative exchanges in 
which a bond swaps partners in a single
concerted move while conserving the total bond number, and so the method does not
capture the limit used for studying vitrimers where relaxation proceeds entirely through such
swaps at fixed connectivity. Adding a swap move would require a three-body
acceptance criterion rather than the pairwise one of \cref{eq:pon}.

Within a sweep, the trials of each reaction are made independent by construction:
proposals are matched so that they act on disjoint pairs, and each reaction reads
occupancies as they stood before it was processed. That independence is what
makes detailed balance follow from the criterion applied to each trial, but it
also means that no site undergoes more than one event per reaction per sweep,
so the trigger period $n$ must be short compared with the bond lifetime
for the prescribed rates to be recovered. Finally, the rate modifiers of
\cref{sec:modifiers} are mutually exclusive, so a bond cannot at present be made
both temperature-dependent and load-dependent.
Other forms of load-dependence such as catch-bonding would require further additions to the code to represent the 
non-monotonic dependence on tension.

By combining a robust scheme for dynamic bonding with a flexible template-based framework driving self-assembly workflows, 
we feel pySNAP is a powerful tool that will rapidly enable studies of a wide range of soft-matter systems.

\begin{acknowledgments}
We thank Kangxin Liu, Tsung Lun Lee, Rakshit Jain, and William Kimball for helping test the software framework and providing valuable feedback.
This work was supported by the National Science Foundation through a DMREF award
to GMH and TMT (grant numbers 2323482 and 2323483). TL was supported by a
fellowship from the Simons Center for Computational Physical Chemistry at NYU
(SCCPC, Simons Foundation Grant MPS-T-MPS-00839534, MET). This work was also supported by the Army Research Office under Grant Number W911NF-23-1-0387. Computational work was
supported in part through the NYU IT High Performance Computing resources,
services, and staff expertise, and simulations were partially executed on
resources purchased by the SCCPC. We acknowledge the Texas Advanced Computing Center (TACC) at The University of Texas at Austin for providing HPC resources.
\end{acknowledgments}

\section*{Data Availability}
The source code of DyBond-v2 and pySNAP is available at
\url{https://github.com/hocky-research-group/DyBond-v2} and
\url{https://github.com/hocky-research-group/pySNAP}.
pySNAP documentation is available at \url{https://hockygroup.com/pySNAP}.
Input files, analysis scripts, and processed data for the three examples of \cref{sec:examples}, together with the notebook that builds \cref{fig:NC,fig:composite,fig:droplet}, are available at \url{https://github.com/hocky-research-group/pySNAP-paper}.

\appendix

\section{Numerical validation of the dynamic bonding implementation}
\label{sec:validation}

The test suite included with the plugin checks the competition between reactions
against an answer computed independently of it, and the tests of the molecule
exclusion are run once per device the host provides, so a divergence between the
CPU and GPU paths in that bookkeeping fails the suite on any machine that has both.

That the scheme reproduces the correct equilibrium is checked exactly in a frozen
geometry. With four reactions $A\!-\!B$, $A\!-\!C$, $D\!-\!B$, and $D\!-\!C$
competing for shared binders $B$ and $C$, every pair held at the bond rest
length, the seven accessible states give the partition function
\begin{equation}
\label{eq:Z}
\begin{split}
Z = 1 &+ K_{AB} + K_{AC} + K_{DB} + K_{DC} \\
      &+ K_{AB}K_{DC} + K_{AC}K_{DB},
\end{split}
\end{equation}
with, for example, $P(A\!-\!B) = K_{AB}(1 + K_{DC})/Z$. Because every candidate
sits at contact, $U_\mathrm{b} = 0$ and the acceptance factor is unity, so what
the comparison isolates is the competition of the four reactions for the shared
sites: the measured populations reproduce these occupancies to within $0.02$ in
probability, and no site is found in more than one bond. Releasing the same four
reactions into a mobile, weakly binding soup turns the test into a dynamic one,
in which bonds exchange partners many times over the course of a single run---the number of
distinct bonded pairs seen exceeds the largest number ever present at once by
more than a factor of three---while the branching still favors the partner with
the larger equilibrium constant and is accurate within fluctuations of the expected result.

The remaining tests cover the bookkeeping on which the independence argument of
\cref{sec:sweep} rests. Unit valence is checked to survive both heavy bond
turnover and HOOMD-blue's particle sorter running throughout, which makes a
particle's tag differ from its position in memory. The molecule exclusion is
checked in both directions, with a control pair that must bond either way.
Two reactions sharing the same pair of endpoint types but differing in
$k_\mathrm{off}$ are checked to unbind by their own bond type rather than by the
type pair. Finally, the zero-temperature limit is checked to be well defined at
both ends, so that neither the acceptance factor nor the slip-bond modifier
produces an undefined rate as $T \to 0$.

\section{Geometric relationship between patch outer radius and droplet diameter}
\label{sec:geometry}

According to the geometry outlined in \cref{fig:droplet}C, the patch outer radius $r$ is given by
\begin{align}
    r&=\sqrt{(R+1.5)^2-(R+\delta-l)^2}.
\end{align}
Here, we omit the unit $\sigma$ from these quantities for simplicity. Since we consider the regime $R\gg1$, we approximate $r$ by keeping only the leading order:
\begin{equation}
    r\simeq\sqrt{1.5+l-\delta}\sqrt{D}.
\end{equation}

By further assuming that $\delta$ is independent of $D$, we predict a linear relationship between patch outer radius $r$ and square root of the droplet diameter $D$.

\section{A rigid-core droplet model}
\label{sec:rigid_core}

\begin{figure}[t]
\includegraphics{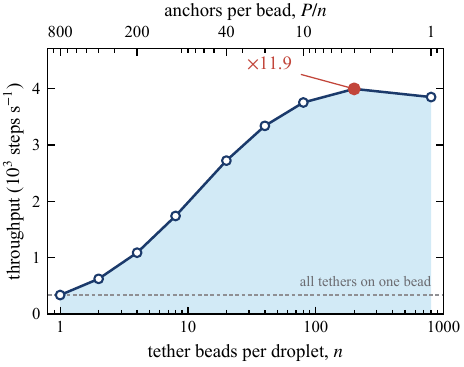}
\caption{Throughput of the rigid-core droplet model against the number $n$ of tether
beads per droplet, for two droplets of diameter $D = 100\sigma$ with $P = 800$ binding
sites each, on one NVIDIA L40S GPU with all pair, bond, and angle forces active. The upper axis gives the number of anchors bonded to
each bead, $P/n$. At $n = 1$ every tether bond of a droplet acts on a single particle, as
in the original model; spreading the bonds over $n = 200$ beads raises the throughput by
a factor of 11.9.}
\label{fig:rigid_core_benchmark}
\end{figure}

In the droplet model as first written, each anchor is bonded to the particle at the center
of its droplet, and that particle carries all $P$ tether bonds of the droplet. The
simulations of \cref{fig:droplet} instead spread those bonds over $n$ tether beads. The
beads sit exactly at the droplet center, coincident with the core particle, and are held
there as constituents of a rigid body whose central particle is the core; each bead is
bonded to $P/n$ of the anchors. Because every bead is at the center, each tether bond spans
the same vector it spanned from the core, with the same harmonic potential, and the
angular potential that keeps a binding site radial, now taken over bead, anchor and
binding site, has its vertex where it was. The forces on anchors and binding sites are
therefore unchanged. The forces on the beads are transferred to the core, which carries
the mass of the whole body as HOOMD-blue's rigid-body integration prescribes, and since
they act at the center they exert no torque. The beads interact with nothing else: their
pair cutoffs are zero, and pairs within a body are excluded from the neighbor list.

\Cref{fig:rigid_core_benchmark} measures the gain for two droplets of diameter
$D = 100\sigma$ with $P = 800$ binding sites each, timed over 2000 steps after a 100-step
warm-up. The throughput rises from 336 time steps per second with a single bead to 3994
with $n = 200$ and levels off beyond, at 3849 with one bead per anchor. The production
runs use $n = P/4$.

We checked the equivalence at the geometry of \cref{fig:droplet}A, $D = 100\sigma$ and
$P = 200$, with $n = 50$, running both models through the same equilibration, placement at
contact, and $250\tau$ of production. The distributions of anchor and binding-site
distances from the droplet center, of the core--anchor--binding-site angle, and of the
nearest-neighbor spacing between anchors differ between the two models by a two-sample
Kolmogorov--Smirnov statistic of at most 0.011, less than the same statistic between the
first and second halves of the original model's own run, at most 0.025, and dynamic bonds
between the droplets accumulate at a similar rate. On the same GPU model, the rigid-core
system ran about six times faster than the original. Each droplet diameter in
\cref{fig:droplet} was then simulated for $2\times10^{4}\tau$ of production.

\bibliography{pySNAP}

\begin{thebibliography}{65}%
\makeatletter
\providecommand \@ifxundefined [1]{%
 \@ifx{#1\undefined}
}%
\providecommand \@ifnum [1]{%
 \ifnum #1\expandafter \@firstoftwo
 \else \expandafter \@secondoftwo
 \fi
}%
\providecommand \@ifx [1]{%
 \ifx #1\expandafter \@firstoftwo
 \else \expandafter \@secondoftwo
 \fi
}%
\providecommand \natexlab [1]{#1}%
\providecommand \enquote  [1]{``#1''}%
\providecommand \bibnamefont  [1]{#1}%
\providecommand \bibfnamefont [1]{#1}%
\providecommand \citenamefont [1]{#1}%
\providecommand \href@noop [0]{\@secondoftwo}%
\providecommand \href [0]{\begingroup \@sanitize@url \@href}%
\providecommand \@href[1]{\@@startlink{#1}\@@href}%
\providecommand \@@href[1]{\endgroup#1\@@endlink}%
\providecommand \@sanitize@url [0]{\catcode `\\12\catcode `\$12\catcode
  `\&12\catcode `\#12\catcode `\^12\catcode `\_12\catcode `\%12\relax}%
\providecommand \@@startlink[1]{}%
\providecommand \@@endlink[0]{}%
\providecommand \url  [0]{\begingroup\@sanitize@url \@url }%
\providecommand \@url [1]{\endgroup\@href {#1}{\urlprefix }}%
\providecommand \urlprefix  [0]{URL }%
\providecommand \Eprint [0]{\href }%
\providecommand \doibase [0]{https://doi.org/}%
\providecommand \selectlanguage [0]{\@gobble}%
\providecommand \bibinfo  [0]{\@secondoftwo}%
\providecommand \bibfield  [0]{\@secondoftwo}%
\providecommand \translation [1]{[#1]}%
\providecommand \BibitemOpen [0]{}%
\providecommand \bibitemStop [0]{}%
\providecommand \bibitemNoStop [0]{.\EOS\space}%
\providecommand \EOS [0]{\spacefactor3000\relax}%
\providecommand \BibitemShut  [1]{\csname bibitem#1\endcsname}%
\let\auto@bib@innerbib\@empty
\bibitem [{\citenamefont {Zhang}\ \emph {et~al.}(2013)\citenamefont {Zhang},
  \citenamefont {Lu}, \citenamefont {Yager}, \citenamefont {Van Der~Lelie},\
  and\ \citenamefont {Gang}}]{zhang2013general}%
  \BibitemOpen
  \bibfield  {author} {\bibinfo {author} {\bibfnamefont {Y.}~\bibnamefont
  {Zhang}}, \bibinfo {author} {\bibfnamefont {F.}~\bibnamefont {Lu}}, \bibinfo
  {author} {\bibfnamefont {K.~G.}\ \bibnamefont {Yager}}, \bibinfo {author}
  {\bibfnamefont {D.}~\bibnamefont {Van Der~Lelie}},\ and\ \bibinfo {author}
  {\bibfnamefont {O.}~\bibnamefont {Gang}},\ }\bibfield  {title} {\bibinfo
  {title} {A general strategy for the {DNA}-mediated self-assembly of
  functional nanoparticles into heterogeneous systems},\ }\href@noop {}
  {\bibfield  {journal} {\bibinfo  {journal} {Nat. Nanotechnol.}\ }\textbf
  {\bibinfo {volume} {8}},\ \bibinfo {pages} {865} (\bibinfo {year}
  {2013})}\BibitemShut {NoStop}%
\bibitem [{\citenamefont {Kay}(2016)}]{kay2016dynamic}%
  \BibitemOpen
  \bibfield  {author} {\bibinfo {author} {\bibfnamefont {E.~R.}\ \bibnamefont
  {Kay}},\ }\bibfield  {title} {\bibinfo {title} {Dynamic covalent nanoparticle
  building blocks},\ }\href@noop {} {\bibfield  {journal} {\bibinfo  {journal}
  {Chem. Eur. J.}\ }\textbf {\bibinfo {volume} {22}},\ \bibinfo {pages} {10706}
  (\bibinfo {year} {2016})}\BibitemShut {NoStop}%
\bibitem [{\citenamefont {Bomboi}\ \emph {et~al.}(2016)\citenamefont {Bomboi},
  \citenamefont {Romano}, \citenamefont {Leo}, \citenamefont
  {Fernandez-Castanon}, \citenamefont {Cerbino}, \citenamefont {Bellini},
  \citenamefont {Bordi}, \citenamefont {Filetici},\ and\ \citenamefont
  {Sciortino}}]{bomboi2016re}%
  \BibitemOpen
  \bibfield  {author} {\bibinfo {author} {\bibfnamefont {F.}~\bibnamefont
  {Bomboi}}, \bibinfo {author} {\bibfnamefont {F.}~\bibnamefont {Romano}},
  \bibinfo {author} {\bibfnamefont {M.}~\bibnamefont {Leo}}, \bibinfo {author}
  {\bibfnamefont {J.}~\bibnamefont {Fernandez-Castanon}}, \bibinfo {author}
  {\bibfnamefont {R.}~\bibnamefont {Cerbino}}, \bibinfo {author} {\bibfnamefont
  {T.}~\bibnamefont {Bellini}}, \bibinfo {author} {\bibfnamefont
  {F.}~\bibnamefont {Bordi}}, \bibinfo {author} {\bibfnamefont
  {P.}~\bibnamefont {Filetici}},\ and\ \bibinfo {author} {\bibfnamefont
  {F.}~\bibnamefont {Sciortino}},\ }\bibfield  {title} {\bibinfo {title}
  {Re-entrant {DNA} gels},\ }\href@noop {} {\bibfield  {journal} {\bibinfo
  {journal} {Nat. Commun.}\ }\textbf {\bibinfo {volume} {7}},\ \bibinfo {pages}
  {1} (\bibinfo {year} {2016})}\BibitemShut {NoStop}%
\bibitem [{\citenamefont {Richardson}\ \emph {et~al.}(2019)\citenamefont
  {Richardson}, \citenamefont {Wilcox}, \citenamefont {Randolph},\ and\
  \citenamefont {Anseth}}]{richardson2019}%
  \BibitemOpen
  \bibfield  {author} {\bibinfo {author} {\bibfnamefont {B.~M.}\ \bibnamefont
  {Richardson}}, \bibinfo {author} {\bibfnamefont {D.~G.}\ \bibnamefont
  {Wilcox}}, \bibinfo {author} {\bibfnamefont {M.~A.}\ \bibnamefont
  {Randolph}},\ and\ \bibinfo {author} {\bibfnamefont {K.~S.}\ \bibnamefont
  {Anseth}},\ }\bibfield  {title} {\bibinfo {title} {Hydrazone covalent
  adaptable networks modulate extracellular matrix deposition for cartilage
  tissue engineering},\ }\href {https://doi.org/10.1016/j.actbio.2018.11.014}
  {\bibfield  {journal} {\bibinfo  {journal} {Acta Biomater.}\ }\textbf
  {\bibinfo {volume} {83}},\ \bibinfo {pages} {71} (\bibinfo {year}
  {2019})}\BibitemShut {NoStop}%
\bibitem [{\citenamefont {Marro}\ \emph {et~al.}(2022)\citenamefont {Marro},
  \citenamefont {Suo}, \citenamefont {Naden},\ and\ \citenamefont
  {Kay}}]{marro2022constitutionally}%
  \BibitemOpen
  \bibfield  {author} {\bibinfo {author} {\bibfnamefont {N.}~\bibnamefont
  {Marro}}, \bibinfo {author} {\bibfnamefont {R.}~\bibnamefont {Suo}}, \bibinfo
  {author} {\bibfnamefont {A.~B.}\ \bibnamefont {Naden}},\ and\ \bibinfo
  {author} {\bibfnamefont {E.~R.}\ \bibnamefont {Kay}},\ }\bibfield  {title}
  {\bibinfo {title} {Constitutionally selective dynamic covalent nanoparticle
  assembly},\ }\href@noop {} {\bibfield  {journal} {\bibinfo  {journal} {J. Am.
  Chem. Soc.}\ }\textbf {\bibinfo {volume} {144}},\ \bibinfo {pages} {14310}
  (\bibinfo {year} {2022})}\BibitemShut {NoStop}%
\bibitem [{\citenamefont {Peng}\ \emph {et~al.}(2020)\citenamefont {Peng},
  \citenamefont {Li}, \citenamefont {Song},\ and\ \citenamefont
  {Yu}}]{peng2020}%
  \BibitemOpen
  \bibfield  {author} {\bibinfo {author} {\bibfnamefont {P.}~\bibnamefont
  {Peng}}, \bibinfo {author} {\bibfnamefont {Y.}~\bibnamefont {Li}}, \bibinfo
  {author} {\bibfnamefont {W.}~\bibnamefont {Song}},\ and\ \bibinfo {author}
  {\bibfnamefont {X.}~\bibnamefont {Yu}},\ }\bibfield  {title} {\bibinfo
  {title} {Self-healing organogels and hydrogels constructed by self-assembled
  bis-terpyridine complex with selective metal ions},\ }\href
  {https://doi.org/10.1016/j.colsurfa.2020.124439} {\bibfield  {journal}
  {\bibinfo  {journal} {Colloids Surf. A}\ }\textbf {\bibinfo {volume} {589}},\
  \bibinfo {pages} {124439} (\bibinfo {year} {2020})}\BibitemShut {NoStop}%
\bibitem [{\citenamefont {Song}\ \emph {et~al.}(2020)\citenamefont {Song},
  \citenamefont {Rizvi}, \citenamefont {Lynch}, \citenamefont {Ilavsky},
  \citenamefont {Mankus}, \citenamefont {Tracy}, \citenamefont {McKinley},\
  and\ \citenamefont {Holten-Andersen}}]{song2020programmable}%
  \BibitemOpen
  \bibfield  {author} {\bibinfo {author} {\bibfnamefont {J.}~\bibnamefont
  {Song}}, \bibinfo {author} {\bibfnamefont {M.~H.}\ \bibnamefont {Rizvi}},
  \bibinfo {author} {\bibfnamefont {B.~B.}\ \bibnamefont {Lynch}}, \bibinfo
  {author} {\bibfnamefont {J.}~\bibnamefont {Ilavsky}}, \bibinfo {author}
  {\bibfnamefont {D.}~\bibnamefont {Mankus}}, \bibinfo {author} {\bibfnamefont
  {J.~B.}\ \bibnamefont {Tracy}}, \bibinfo {author} {\bibfnamefont {G.~H.}\
  \bibnamefont {McKinley}},\ and\ \bibinfo {author} {\bibfnamefont
  {N.}~\bibnamefont {Holten-Andersen}},\ }\bibfield  {title} {\bibinfo {title}
  {Programmable anisotropy and percolation in supramolecular patchy particle
  gels},\ }\href@noop {} {\bibfield  {journal} {\bibinfo  {journal} {ACS Nano}\
  }\textbf {\bibinfo {volume} {14}},\ \bibinfo {pages} {17018} (\bibinfo {year}
  {2020})}\BibitemShut {NoStop}%
\bibitem [{\citenamefont {Dominguez}\ \emph {et~al.}(2020)\citenamefont
  {Dominguez}, \citenamefont {Howard}, \citenamefont {Maier}, \citenamefont
  {Valenzuela}, \citenamefont {Sherman}, \citenamefont {Reuther}, \citenamefont
  {Reimnitz}, \citenamefont {Kang}, \citenamefont {Cho}, \citenamefont {Gibbs}
  \emph {et~al.}}]{dominguez2020assembly}%
  \BibitemOpen
  \bibfield  {author} {\bibinfo {author} {\bibfnamefont {M.~N.}\ \bibnamefont
  {Dominguez}}, \bibinfo {author} {\bibfnamefont {M.~P.}\ \bibnamefont
  {Howard}}, \bibinfo {author} {\bibfnamefont {J.~M.}\ \bibnamefont {Maier}},
  \bibinfo {author} {\bibfnamefont {S.~A.}\ \bibnamefont {Valenzuela}},
  \bibinfo {author} {\bibfnamefont {Z.~M.}\ \bibnamefont {Sherman}}, \bibinfo
  {author} {\bibfnamefont {J.~F.}\ \bibnamefont {Reuther}}, \bibinfo {author}
  {\bibfnamefont {L.~C.}\ \bibnamefont {Reimnitz}}, \bibinfo {author}
  {\bibfnamefont {J.}~\bibnamefont {Kang}}, \bibinfo {author} {\bibfnamefont
  {S.~H.}\ \bibnamefont {Cho}}, \bibinfo {author} {\bibfnamefont {S.~L.}\
  \bibnamefont {Gibbs}}, \emph {et~al.},\ }\bibfield  {title} {\bibinfo {title}
  {Assembly of linked nanocrystal colloids by reversible covalent bonds},\
  }\href@noop {} {\bibfield  {journal} {\bibinfo  {journal} {Chem. Mater.}\
  }\textbf {\bibinfo {volume} {32}},\ \bibinfo {pages} {10235} (\bibinfo {year}
  {2020})}\BibitemShut {NoStop}%
\bibitem [{\citenamefont {Kang}\ \emph {et~al.}(2022)\citenamefont {Kang},
  \citenamefont {Valenzuela}, \citenamefont {Lin}, \citenamefont {Dominguez},
  \citenamefont {Sherman}, \citenamefont {Truskett}, \citenamefont {Anslyn},\
  and\ \citenamefont {Milliron}}]{kang2022colorimetric}%
  \BibitemOpen
  \bibfield  {author} {\bibinfo {author} {\bibfnamefont {J.}~\bibnamefont
  {Kang}}, \bibinfo {author} {\bibfnamefont {S.~A.}\ \bibnamefont
  {Valenzuela}}, \bibinfo {author} {\bibfnamefont {E.~Y.}\ \bibnamefont {Lin}},
  \bibinfo {author} {\bibfnamefont {M.~N.}\ \bibnamefont {Dominguez}}, \bibinfo
  {author} {\bibfnamefont {Z.~M.}\ \bibnamefont {Sherman}}, \bibinfo {author}
  {\bibfnamefont {T.~M.}\ \bibnamefont {Truskett}}, \bibinfo {author}
  {\bibfnamefont {E.~V.}\ \bibnamefont {Anslyn}},\ and\ \bibinfo {author}
  {\bibfnamefont {D.~J.}\ \bibnamefont {Milliron}},\ }\bibfield  {title}
  {\bibinfo {title} {Colorimetric quantification of linking in thermoreversible
  nanocrystal gel assemblies},\ }\href@noop {} {\bibfield  {journal} {\bibinfo
  {journal} {Sci. Adv.}\ }\textbf {\bibinfo {volume} {8}},\ \bibinfo {pages}
  {eabm7364} (\bibinfo {year} {2022})}\BibitemShut {NoStop}%
\bibitem [{\citenamefont {Kang}\ \emph {et~al.}(2025)\citenamefont {Kang},
  \citenamefont {Qian}, \citenamefont {Lee}, \citenamefont {Conrad},
  \citenamefont {Oberlander}, \citenamefont {Berry}, \citenamefont {Liu},
  \citenamefont {Anslyn}, \citenamefont {Truskett},\ and\ \citenamefont
  {Milliron}}]{kang2025colloidal}%
  \BibitemOpen
  \bibfield  {author} {\bibinfo {author} {\bibfnamefont {J.}~\bibnamefont
  {Kang}}, \bibinfo {author} {\bibfnamefont {D.}~\bibnamefont {Qian}}, \bibinfo
  {author} {\bibfnamefont {J.}~\bibnamefont {Lee}}, \bibinfo {author}
  {\bibfnamefont {D.}~\bibnamefont {Conrad}}, \bibinfo {author} {\bibfnamefont
  {J.}~\bibnamefont {Oberlander}}, \bibinfo {author} {\bibfnamefont {M.~W.}\
  \bibnamefont {Berry}}, \bibinfo {author} {\bibfnamefont {J.}~\bibnamefont
  {Liu}}, \bibinfo {author} {\bibfnamefont {E.}~\bibnamefont {Anslyn}},
  \bibinfo {author} {\bibfnamefont {T.}~\bibnamefont {Truskett}},\ and\
  \bibinfo {author} {\bibfnamefont {D.}~\bibnamefont {Milliron}},\ }\bibfield
  {title} {\bibinfo {title} {Colloidal phase control in plasmonic metal oxide
  nanocrystals via competitive metal--ligand equilibria},\ }\href@noop {}
  {\bibfield  {journal} {\bibinfo  {journal} {Angew. Chem. Int. Ed.}\ }\textbf
  {\bibinfo {volume} {64}},\ \bibinfo {pages} {e18965} (\bibinfo {year}
  {2025})}\BibitemShut {NoStop}%
\bibitem [{\citenamefont {Kim}\ \emph {et~al.}(2025)\citenamefont {Kim},
  \citenamefont {Livermore}, \citenamefont {Rowan},\ and\ \citenamefont
  {Jaeger}}]{kim2025dense}%
  \BibitemOpen
  \bibfield  {author} {\bibinfo {author} {\bibfnamefont {H.}~\bibnamefont
  {Kim}}, \bibinfo {author} {\bibfnamefont {S.~M.}\ \bibnamefont {Livermore}},
  \bibinfo {author} {\bibfnamefont {S.~J.}\ \bibnamefont {Rowan}},\ and\
  \bibinfo {author} {\bibfnamefont {H.~M.}\ \bibnamefont {Jaeger}},\ }\bibfield
   {title} {\bibinfo {title} {Dense suspensions as trainable rheological
  metafluids},\ }\href@noop {} {\bibfield  {journal} {\bibinfo  {journal}
  {Proc. Natl. Acad. Sci. USA}\ }\textbf {\bibinfo {volume} {122}},\ \bibinfo
  {pages} {e2509525122} (\bibinfo {year} {2025})}\BibitemShut {NoStop}%
\bibitem [{\citenamefont {Bertsch}\ \emph {et~al.}(2023)\citenamefont
  {Bertsch}, \citenamefont {Diba}, \citenamefont {Mooney},\ and\ \citenamefont
  {Leeuwenburgh}}]{Bertsch2023Self-HealingRegeneration}%
  \BibitemOpen
  \bibfield  {author} {\bibinfo {author} {\bibfnamefont {P.}~\bibnamefont
  {Bertsch}}, \bibinfo {author} {\bibfnamefont {M.}~\bibnamefont {Diba}},
  \bibinfo {author} {\bibfnamefont {D.~J.}\ \bibnamefont {Mooney}},\ and\
  \bibinfo {author} {\bibfnamefont {S.~C.~G.}\ \bibnamefont {Leeuwenburgh}},\
  }\bibfield  {title} {\bibinfo {title} {Self-healing injectable hydrogels for
  tissue regeneration},\ }\href {https://doi.org/10.1021/acs.chemrev.2c00179}
  {\bibfield  {journal} {\bibinfo  {journal} {Chem. Rev.}\ }\textbf {\bibinfo
  {volume} {123}},\ \bibinfo {pages} {834} (\bibinfo {year}
  {2023})}\BibitemShut {NoStop}%
\bibitem [{\citenamefont {Wang}\ and\ \citenamefont
  {Urban}(2020)}]{wang2020self}%
  \BibitemOpen
  \bibfield  {author} {\bibinfo {author} {\bibfnamefont {S.}~\bibnamefont
  {Wang}}\ and\ \bibinfo {author} {\bibfnamefont {M.~W.}\ \bibnamefont
  {Urban}},\ }\bibfield  {title} {\bibinfo {title} {Self-healing polymers},\
  }\href@noop {} {\bibfield  {journal} {\bibinfo  {journal} {Nat. Rev. Mater.}\
  }\textbf {\bibinfo {volume} {5}},\ \bibinfo {pages} {562} (\bibinfo {year}
  {2020})}\BibitemShut {NoStop}%
\bibitem [{\citenamefont {Rosales}\ and\ \citenamefont
  {Anseth}(2016)}]{rosales2016design}%
  \BibitemOpen
  \bibfield  {author} {\bibinfo {author} {\bibfnamefont {A.~M.}\ \bibnamefont
  {Rosales}}\ and\ \bibinfo {author} {\bibfnamefont {K.~S.}\ \bibnamefont
  {Anseth}},\ }\bibfield  {title} {\bibinfo {title} {The design of reversible
  hydrogels to capture extracellular matrix dynamics},\ }\href@noop {}
  {\bibfield  {journal} {\bibinfo  {journal} {Nat. Rev. Mater.}\ }\textbf
  {\bibinfo {volume} {1}},\ \bibinfo {pages} {15012} (\bibinfo {year}
  {2016})}\BibitemShut {NoStop}%
\bibitem [{\citenamefont {Loebel}\ \emph {et~al.}(2017)\citenamefont {Loebel},
  \citenamefont {Rodell}, \citenamefont {Chen},\ and\ \citenamefont
  {Burdick}}]{loebel2017shear}%
  \BibitemOpen
  \bibfield  {author} {\bibinfo {author} {\bibfnamefont {C.}~\bibnamefont
  {Loebel}}, \bibinfo {author} {\bibfnamefont {C.~B.}\ \bibnamefont {Rodell}},
  \bibinfo {author} {\bibfnamefont {M.~H.}\ \bibnamefont {Chen}},\ and\
  \bibinfo {author} {\bibfnamefont {J.~A.}\ \bibnamefont {Burdick}},\
  }\bibfield  {title} {\bibinfo {title} {Shear-thinning and self-healing
  hydrogels as injectable therapeutics and for {3D-printing}},\ }\href@noop {}
  {\bibfield  {journal} {\bibinfo  {journal} {Nat. Protoc.}\ }\textbf {\bibinfo
  {volume} {12}},\ \bibinfo {pages} {1521} (\bibinfo {year}
  {2017})}\BibitemShut {NoStop}%
\bibitem [{\citenamefont {Kloxin}\ and\ \citenamefont
  {Bowman}(2013)}]{Kloxin2013CovalentSystems}%
  \BibitemOpen
  \bibfield  {author} {\bibinfo {author} {\bibfnamefont {C.~J.}\ \bibnamefont
  {Kloxin}}\ and\ \bibinfo {author} {\bibfnamefont {C.~N.}\ \bibnamefont
  {Bowman}},\ }\bibfield  {title} {\bibinfo {title} {{Covalent adaptable
  networks: Smart, reconfigurable and responsive network systems}},\ }\href
  {https://doi.org/10.1039/c3cs60046g} {\bibfield  {journal} {\bibinfo
  {journal} {Chem. Soc. Rev.}\ }\textbf {\bibinfo {volume} {42}},\ \bibinfo
  {pages} {7161} (\bibinfo {year} {2013})}\BibitemShut {NoStop}%
\bibitem [{\citenamefont {Webber}\ and\ \citenamefont
  {Tibbitt}(2022)}]{Webber2022DynamicInteractions}%
  \BibitemOpen
  \bibfield  {author} {\bibinfo {author} {\bibfnamefont {M.~J.}\ \bibnamefont
  {Webber}}\ and\ \bibinfo {author} {\bibfnamefont {M.~W.}\ \bibnamefont
  {Tibbitt}},\ }\bibfield  {title} {\bibinfo {title} {{Dynamic and
  reconfigurable materials from reversible network interactions}},\ }\href
  {https://doi.org/10.1038/s41578-021-00412-x} {\bibfield  {journal} {\bibinfo
  {journal} {Nat. Rev. Mater.}\ }\textbf {\bibinfo {volume} {7}},\ \bibinfo
  {pages} {541} (\bibinfo {year} {2022})}\BibitemShut {NoStop}%
\bibitem [{\citenamefont {Karatrantos}\ \emph {et~al.}(2024)\citenamefont
  {Karatrantos}, \citenamefont {Couture}, \citenamefont {Hesse},\ and\
  \citenamefont {Schmidt}}]{Karatrantos2024MolecularReview}%
  \BibitemOpen
  \bibfield  {author} {\bibinfo {author} {\bibfnamefont {A.~V.}\ \bibnamefont
  {Karatrantos}}, \bibinfo {author} {\bibfnamefont {O.}~\bibnamefont
  {Couture}}, \bibinfo {author} {\bibfnamefont {C.}~\bibnamefont {Hesse}},\
  and\ \bibinfo {author} {\bibfnamefont {D.~F.}\ \bibnamefont {Schmidt}},\
  }\bibfield  {title} {\bibinfo {title} {Molecular simulation of covalent
  adaptable networks and vitrimers: A review},\ }\bibfield  {journal} {\bibinfo
   {journal} {Polymers}\ }\textbf {\bibinfo {volume} {16}},\ \href
  {https://doi.org/10.3390/polym16101373} {10.3390/polym16101373} (\bibinfo
  {year} {2024})\BibitemShut {NoStop}%
\bibitem [{\citenamefont {Sun}\ \emph {et~al.}(2023)\citenamefont {Sun},
  \citenamefont {Wan}, \citenamefont {Shen}, \citenamefont {He}, \citenamefont
  {Zhou}, \citenamefont {Wang}, \citenamefont {Yang},\ and\ \citenamefont
  {Shi}}]{sun2023modeling}%
  \BibitemOpen
  \bibfield  {author} {\bibinfo {author} {\bibfnamefont {Y.}~\bibnamefont
  {Sun}}, \bibinfo {author} {\bibfnamefont {K.}~\bibnamefont {Wan}}, \bibinfo
  {author} {\bibfnamefont {W.}~\bibnamefont {Shen}}, \bibinfo {author}
  {\bibfnamefont {J.}~\bibnamefont {He}}, \bibinfo {author} {\bibfnamefont
  {T.}~\bibnamefont {Zhou}}, \bibinfo {author} {\bibfnamefont {H.}~\bibnamefont
  {Wang}}, \bibinfo {author} {\bibfnamefont {H.}~\bibnamefont {Yang}},\ and\
  \bibinfo {author} {\bibfnamefont {X.}~\bibnamefont {Shi}},\ }\bibfield
  {title} {\bibinfo {title} {Modeling exchange reactions in covalent adaptable
  networks with machine learning force fields},\ }\href@noop {} {\bibfield
  {journal} {\bibinfo  {journal} {Macromolecules}\ }\textbf {\bibinfo {volume}
  {56}},\ \bibinfo {pages} {9003} (\bibinfo {year} {2023})}\BibitemShut
  {NoStop}%
\bibitem [{\citenamefont {Akhtar}\ \emph {et~al.}(2026)\citenamefont {Akhtar},
  \citenamefont {Kunnikuruvan}, \citenamefont {Yadav},\ and\ \citenamefont
  {Patra}}]{akhtar2026molecular}%
  \BibitemOpen
  \bibfield  {author} {\bibinfo {author} {\bibfnamefont {J.}~\bibnamefont
  {Akhtar}}, \bibinfo {author} {\bibfnamefont {S.}~\bibnamefont
  {Kunnikuruvan}}, \bibinfo {author} {\bibfnamefont {S.~K.}\ \bibnamefont
  {Yadav}},\ and\ \bibinfo {author} {\bibfnamefont {T.~K.}\ \bibnamefont
  {Patra}},\ }\bibfield  {title} {\bibinfo {title} {Molecular origins of
  dynamic covalent bonds in polymers},\ }\bibfield  {journal} {\bibinfo
  {journal} {ChemRxiv}\ }\href {https://doi.org/10.26434/chemrxiv.15007032/v1}
  {10.26434/chemrxiv.15007032/v1} (\bibinfo {year} {2026})\BibitemShut
  {NoStop}%
\bibitem [{\citenamefont {Ciarella}\ \emph {et~al.}(2018)\citenamefont
  {Ciarella}, \citenamefont {Sciortino},\ and\ \citenamefont
  {Ellenbroek}}]{Ciarella2018DynamicsRelaxation}%
  \BibitemOpen
  \bibfield  {author} {\bibinfo {author} {\bibfnamefont {S.}~\bibnamefont
  {Ciarella}}, \bibinfo {author} {\bibfnamefont {F.}~\bibnamefont
  {Sciortino}},\ and\ \bibinfo {author} {\bibfnamefont {W.~G.}\ \bibnamefont
  {Ellenbroek}},\ }\bibfield  {title} {\bibinfo {title} {{Dynamics of
  Vitrimers: Defects as a Highway to Stress Relaxation}},\ }\bibfield
  {journal} {\bibinfo  {journal} {Phys. Rev. Lett.}\ }\textbf {\bibinfo
  {volume} {121}},\ \href {https://doi.org/10.1103/PhysRevLett.121.058003}
  {10.1103/PhysRevLett.121.058003} (\bibinfo {year} {2018})\BibitemShut
  {NoStop}%
\bibitem [{\citenamefont {Gissinger}\ \emph {et~al.}(2017)\citenamefont
  {Gissinger}, \citenamefont {Jensen},\ and\ \citenamefont
  {Wise}}]{Gissinger2017ModelingSimulations}%
  \BibitemOpen
  \bibfield  {author} {\bibinfo {author} {\bibfnamefont {J.~R.}\ \bibnamefont
  {Gissinger}}, \bibinfo {author} {\bibfnamefont {B.~D.}\ \bibnamefont
  {Jensen}},\ and\ \bibinfo {author} {\bibfnamefont {K.~E.}\ \bibnamefont
  {Wise}},\ }\bibfield  {title} {\bibinfo {title} {{Modeling chemical reactions
  in classical molecular dynamics simulations}},\ }\href
  {https://doi.org/10.1016/j.polymer.2017.09.038} {\bibfield  {journal}
  {\bibinfo  {journal} {Polymer}\ }\textbf {\bibinfo {volume} {128}},\ \bibinfo
  {pages} {211} (\bibinfo {year} {2017})}\BibitemShut {NoStop}%
\bibitem [{\citenamefont {Vashisth}\ \emph {et~al.}(2018)\citenamefont
  {Vashisth}, \citenamefont {Ashraf}, \citenamefont {Zhang}, \citenamefont
  {Bakis},\ and\ \citenamefont {van Duin}}]{Vashisth2018AcceleratedPolymers}%
  \BibitemOpen
  \bibfield  {author} {\bibinfo {author} {\bibfnamefont {A.}~\bibnamefont
  {Vashisth}}, \bibinfo {author} {\bibfnamefont {C.}~\bibnamefont {Ashraf}},
  \bibinfo {author} {\bibfnamefont {W.}~\bibnamefont {Zhang}}, \bibinfo
  {author} {\bibfnamefont {C.~E.}\ \bibnamefont {Bakis}},\ and\ \bibinfo
  {author} {\bibfnamefont {A.~C.~T.}\ \bibnamefont {van Duin}},\ }\bibfield
  {title} {\bibinfo {title} {{Accelerated ReaxFF Simulations for Describing the
  Reactive Cross-Linking of Polymers}},\ }\href
  {https://doi.org/10.1021/ACS.JPCA.8B03826} {\bibfield  {journal} {\bibinfo
  {journal} {J. Phys. Chem. A}\ }\textbf {\bibinfo {volume} {122}},\ \bibinfo
  {pages} {6633} (\bibinfo {year} {2018})}\BibitemShut {NoStop}%
\bibitem [{\citenamefont {Blanco}\ and\ \citenamefont
  {Ko{\v{s}}ovan}(2024)}]{Blanco2024TheMethod}%
  \BibitemOpen
  \bibfield  {author} {\bibinfo {author} {\bibfnamefont {P.~M.}\ \bibnamefont
  {Blanco}}\ and\ \bibinfo {author} {\bibfnamefont {P.}~\bibnamefont
  {Ko{\v{s}}ovan}},\ }\bibfield  {title} {\bibinfo {title} {{The explicit
  bonding reaction ensemble Monte Carlo method}},\ }\bibfield  {journal}
  {\bibinfo  {journal} {J. Comp. Phys.}\ }\textbf {\bibinfo {volume} {161}},\
  \href {https://doi.org/10.1063/5.0226122} {10.1063/5.0226122} (\bibinfo
  {year} {2024})\BibitemShut {NoStop}%
\bibitem [{\citenamefont {Rao}\ \emph {et~al.}(2024)\citenamefont {Rao},
  \citenamefont {Xia},\ and\ \citenamefont {Ni}}]{Rao2024APolymers}%
  \BibitemOpen
  \bibfield  {author} {\bibinfo {author} {\bibfnamefont {P.}~\bibnamefont
  {Rao}}, \bibinfo {author} {\bibfnamefont {X.}~\bibnamefont {Xia}},\ and\
  \bibinfo {author} {\bibfnamefont {R.}~\bibnamefont {Ni}},\ }\bibfield
  {title} {\bibinfo {title} {{A bond swap algorithm for simulating dynamically
  crosslinked polymers}},\ }\href {https://doi.org/10.1063/5.0186553}
  {\bibfield  {journal} {\bibinfo  {journal} {J. Chem. Phys.}\ }\textbf
  {\bibinfo {volume} {160}},\ \bibinfo {pages} {061102} (\bibinfo {year}
  {2024})}\BibitemShut {NoStop}%
\bibitem [{\citenamefont {Mitra}\ \emph {et~al.}(2023)\citenamefont {Mitra},
  \citenamefont {Chang}, \citenamefont {McMullen}, \citenamefont {Puchall},
  \citenamefont {Brujic},\ and\ \citenamefont {Hocky}}]{mitra2023coarse}%
  \BibitemOpen
  \bibfield  {author} {\bibinfo {author} {\bibfnamefont {G.}~\bibnamefont
  {Mitra}}, \bibinfo {author} {\bibfnamefont {C.}~\bibnamefont {Chang}},
  \bibinfo {author} {\bibfnamefont {A.}~\bibnamefont {McMullen}}, \bibinfo
  {author} {\bibfnamefont {D.}~\bibnamefont {Puchall}}, \bibinfo {author}
  {\bibfnamefont {J.}~\bibnamefont {Brujic}},\ and\ \bibinfo {author}
  {\bibfnamefont {G.~M.}\ \bibnamefont {Hocky}},\ }\bibfield  {title} {\bibinfo
  {title} {A coarse-grained simulation model for colloidal self-assembly via
  explicit mobile binders},\ }\href@noop {} {\bibfield  {journal} {\bibinfo
  {journal} {Soft Matter}\ }\textbf {\bibinfo {volume} {19}},\ \bibinfo {pages}
  {4223} (\bibinfo {year} {2023})}\BibitemShut {NoStop}%
\bibitem [{\citenamefont {Jedlinska}\ and\ \citenamefont
  {Riggleman}(2024)}]{Jedlinska2024EffectsCoacervates}%
  \BibitemOpen
  \bibfield  {author} {\bibinfo {author} {\bibfnamefont {Z.~M.}\ \bibnamefont
  {Jedlinska}}\ and\ \bibinfo {author} {\bibfnamefont {R.~A.}\ \bibnamefont
  {Riggleman}},\ }\bibfield  {title} {\bibinfo {title} {{Effects of Associative
  Interactions on the Phase Behavior of Complex Coacervates}},\ }\href
  {https://doi.org/10.1021/acs.macromol.4c00367} {\bibfield  {journal}
  {\bibinfo  {journal} {Macromolecules}\ }\textbf {\bibinfo {volume} {57}},\
  \bibinfo {pages} {4323} (\bibinfo {year} {2024})}\BibitemShut {NoStop}%
\bibitem [{\citenamefont {Freedman}\ \emph {et~al.}(2017)\citenamefont
  {Freedman}, \citenamefont {Banerjee}, \citenamefont {Hocky},\ and\
  \citenamefont {Dinner}}]{Freedman2017ANetworks}%
  \BibitemOpen
  \bibfield  {author} {\bibinfo {author} {\bibfnamefont {S.~L.}\ \bibnamefont
  {Freedman}}, \bibinfo {author} {\bibfnamefont {S.}~\bibnamefont {Banerjee}},
  \bibinfo {author} {\bibfnamefont {G.~M.}\ \bibnamefont {Hocky}},\ and\
  \bibinfo {author} {\bibfnamefont {A.~R.}\ \bibnamefont {Dinner}},\ }\bibfield
   {title} {\bibinfo {title} {{A Versatile Framework for Simulating the Dynamic
  Mechanical Structure of Cytoskeletal Networks}},\ }\href
  {https://doi.org/10.1016/J.BPJ.2017.06.003} {\bibfield  {journal} {\bibinfo
  {journal} {Biophys. J.}\ }\textbf {\bibinfo {volume} {113}},\ \bibinfo
  {pages} {448} (\bibinfo {year} {2017})}\BibitemShut {NoStop}%
\bibitem [{\citenamefont {Lugo}\ \emph {et~al.}(2023)\citenamefont {Lugo},
  \citenamefont {Saikia},\ and\ \citenamefont {Nedelec}}]{Lugo2023ASystems}%
  \BibitemOpen
  \bibfield  {author} {\bibinfo {author} {\bibfnamefont {C.~A.}\ \bibnamefont
  {Lugo}}, \bibinfo {author} {\bibfnamefont {E.}~\bibnamefont {Saikia}},\ and\
  \bibinfo {author} {\bibfnamefont {F.}~\bibnamefont {Nedelec}},\ }\bibfield
  {title} {\bibinfo {title} {{A Typical Workflow to Simulate Cytoskeletal
  Systems}},\ }\href {https://doi.org/10.3791/64125} {\bibfield  {journal}
  {\bibinfo  {journal} {J. Vis. Exp.}\ }\textbf {\bibinfo {volume} {2023}},\
  \bibinfo {pages} {e64125} (\bibinfo {year} {2023})}\BibitemShut {NoStop}%
\bibitem [{\citenamefont {Yan}\ \emph {et~al.}(2022)\citenamefont {Yan},
  \citenamefont {Ansari}, \citenamefont {Lamson}, \citenamefont {Glaser},
  \citenamefont {Blackwell}, \citenamefont {Betterton},\ and\ \citenamefont
  {Shelley}}]{Yan2022TowardAssemblies}%
  \BibitemOpen
  \bibfield  {author} {\bibinfo {author} {\bibfnamefont {W.}~\bibnamefont
  {Yan}}, \bibinfo {author} {\bibfnamefont {S.}~\bibnamefont {Ansari}},
  \bibinfo {author} {\bibfnamefont {A.}~\bibnamefont {Lamson}}, \bibinfo
  {author} {\bibfnamefont {M.~A.}\ \bibnamefont {Glaser}}, \bibinfo {author}
  {\bibfnamefont {R.}~\bibnamefont {Blackwell}}, \bibinfo {author}
  {\bibfnamefont {M.~D.}\ \bibnamefont {Betterton}},\ and\ \bibinfo {author}
  {\bibfnamefont {M.}~\bibnamefont {Shelley}},\ }\bibfield  {title} {\bibinfo
  {title} {{Toward the cellular-scale simulation of motor-driven cytoskeletal
  assemblies}},\ }\href {https://doi.org/10.7554/eLife.74160} {\bibfield
  {journal} {\bibinfo  {journal} {eLife}\ }\textbf {\bibinfo {volume} {11}},\
  \bibinfo {pages} {e74160} (\bibinfo {year} {2022})}\BibitemShut {NoStop}%
\bibitem [{\citenamefont {Popov}\ \emph {et~al.}(2016)\citenamefont {Popov},
  \citenamefont {Komianos},\ and\ \citenamefont
  {Papoian}}]{Popov2016MEDYAN:Networks}%
  \BibitemOpen
  \bibfield  {author} {\bibinfo {author} {\bibfnamefont {K.}~\bibnamefont
  {Popov}}, \bibinfo {author} {\bibfnamefont {J.}~\bibnamefont {Komianos}},\
  and\ \bibinfo {author} {\bibfnamefont {G.~A.}\ \bibnamefont {Papoian}},\
  }\bibfield  {title} {\bibinfo {title} {{MEDYAN: Mechanochemical Simulations
  of Contraction and Polarity Alignment in Actomyosin Networks}},\ }\href
  {https://doi.org/10.1371/journal.pcbi.1004877} {\bibfield  {journal}
  {\bibinfo  {journal} {PLoS Comput. Biol.}\ }\textbf {\bibinfo {volume}
  {12}},\ \bibinfo {pages} {e1004877} (\bibinfo {year} {2016})}\BibitemShut
  {NoStop}%
\bibitem [{\citenamefont {Liu}\ and\ \citenamefont
  {O’Connor}(2024)}]{Liu2024AEquilibrium}%
  \BibitemOpen
  \bibfield  {author} {\bibinfo {author} {\bibfnamefont {S.}~\bibnamefont
  {Liu}}\ and\ \bibinfo {author} {\bibfnamefont {T.~C.}\ \bibnamefont
  {O’Connor}},\ }\bibfield  {title} {\bibinfo {title} {{A Reactive
  Bead–Spring Model for Associative Polymer Melts In and Out of
  Equilibrium}},\ }\href {https://doi.org/10.1021/acs.macromol.3c02022}
  {\bibfield  {journal} {\bibinfo  {journal} {Macromolecules}\ }\textbf
  {\bibinfo {volume} {57}},\ \bibinfo {pages} {1403} (\bibinfo {year}
  {2024})}\BibitemShut {NoStop}%
\bibitem [{\citenamefont {Holoman}\ \emph
  {et~al.}(2026{\natexlab{a}})\citenamefont {Holoman}, \citenamefont {Prajwal},
  \citenamefont {Hocky},\ and\ \citenamefont
  {Truskett}}]{holoman2026simulating}%
  \BibitemOpen
  \bibfield  {author} {\bibinfo {author} {\bibfnamefont {T.~R.}\ \bibnamefont
  {Holoman}}, \bibinfo {author} {\bibfnamefont {B.}~\bibnamefont {Prajwal}},
  \bibinfo {author} {\bibfnamefont {G.~M.}\ \bibnamefont {Hocky}},\ and\
  \bibinfo {author} {\bibfnamefont {T.~M.}\ \bibnamefont {Truskett}},\
  }\bibfield  {title} {\bibinfo {title} {Simulating dynamic bonding in soft
  materials},\ }\href@noop {} {\bibfield  {journal} {\bibinfo  {journal} {Curr.
  Opin. Colloid Interface Sci.}\ }\textbf {\bibinfo {volume} {83}},\ \bibinfo
  {pages} {102019} (\bibinfo {year} {2026}{\natexlab{a}})}\BibitemShut
  {NoStop}%
\bibitem [{\citenamefont {Shu}\ \emph {et~al.}(2024)\citenamefont {Shu},
  \citenamefont {Mitra}, \citenamefont {Alberts}, \citenamefont {Viana},
  \citenamefont {Levy}, \citenamefont {Hocky},\ and\ \citenamefont
  {Holt}}]{shu2024mesoscale}%
  \BibitemOpen
  \bibfield  {author} {\bibinfo {author} {\bibfnamefont {T.}~\bibnamefont
  {Shu}}, \bibinfo {author} {\bibfnamefont {G.}~\bibnamefont {Mitra}}, \bibinfo
  {author} {\bibfnamefont {J.}~\bibnamefont {Alberts}}, \bibinfo {author}
  {\bibfnamefont {M.~P.}\ \bibnamefont {Viana}}, \bibinfo {author}
  {\bibfnamefont {E.~D.}\ \bibnamefont {Levy}}, \bibinfo {author}
  {\bibfnamefont {G.~M.}\ \bibnamefont {Hocky}},\ and\ \bibinfo {author}
  {\bibfnamefont {L.~J.}\ \bibnamefont {Holt}},\ }\bibfield  {title} {\bibinfo
  {title} {Mesoscale molecular assembly is favored by the active, crowded
  cytoplasm},\ }\href@noop {} {\bibfield  {journal} {\bibinfo  {journal} {PRX
  Life}\ }\textbf {\bibinfo {volume} {2}},\ \bibinfo {pages} {033001} (\bibinfo
  {year} {2024})}\BibitemShut {NoStop}%
\bibitem [{\citenamefont {Holoman}\ \emph
  {et~al.}(2026{\natexlab{b}})\citenamefont {Holoman}, \citenamefont {Petix},
  \citenamefont {Howard},\ and\ \citenamefont {Truskett}}]{holoman2026coarse}%
  \BibitemOpen
  \bibfield  {author} {\bibinfo {author} {\bibfnamefont {T.~R.}\ \bibnamefont
  {Holoman}}, \bibinfo {author} {\bibfnamefont {C.~L.}\ \bibnamefont {Petix}},
  \bibinfo {author} {\bibfnamefont {M.~P.}\ \bibnamefont {Howard}},\ and\
  \bibinfo {author} {\bibfnamefont {T.~M.}\ \bibnamefont {Truskett}},\
  }\href@noop {} {\bibinfo {title} {Coarse-graining to create minimalist models
  for dynamic, end-linked star-polymer networks}} (\bibinfo {year}
  {2026}{\natexlab{b}}),\ \Eprint {https://arxiv.org/abs/2609.09594}
  {arXiv:2609.09594} \BibitemShut {NoStop}%
\bibitem [{\citenamefont {Sherman}\ \emph {et~al.}(2023)\citenamefont
  {Sherman}, \citenamefont {Kim}, \citenamefont {Kang}, \citenamefont {Roman},
  \citenamefont {Crory}, \citenamefont {Conrad}, \citenamefont {Valenzuela},
  \citenamefont {Lin}, \citenamefont {Dominguez}, \citenamefont {Gibbs},
  \citenamefont {Anslyn}, \citenamefont {Milliron},\ and\ \citenamefont
  {Truskett}}]{Sherman2023}%
  \BibitemOpen
  \bibfield  {author} {\bibinfo {author} {\bibfnamefont {Z.~M.}\ \bibnamefont
  {Sherman}}, \bibinfo {author} {\bibfnamefont {K.}~\bibnamefont {Kim}},
  \bibinfo {author} {\bibfnamefont {J.}~\bibnamefont {Kang}}, \bibinfo {author}
  {\bibfnamefont {B.~J.}\ \bibnamefont {Roman}}, \bibinfo {author}
  {\bibfnamefont {H.~S.~N.}\ \bibnamefont {Crory}}, \bibinfo {author}
  {\bibfnamefont {D.~L.}\ \bibnamefont {Conrad}}, \bibinfo {author}
  {\bibfnamefont {S.~A.}\ \bibnamefont {Valenzuela}}, \bibinfo {author}
  {\bibfnamefont {E.~Y.}\ \bibnamefont {Lin}}, \bibinfo {author} {\bibfnamefont
  {M.~N.}\ \bibnamefont {Dominguez}}, \bibinfo {author} {\bibfnamefont {S.~L.}\
  \bibnamefont {Gibbs}}, \bibinfo {author} {\bibfnamefont {E.~V.}\ \bibnamefont
  {Anslyn}}, \bibinfo {author} {\bibfnamefont {D.~J.}\ \bibnamefont
  {Milliron}},\ and\ \bibinfo {author} {\bibfnamefont {T.~M.}\ \bibnamefont
  {Truskett}},\ }\bibfield  {title} {\bibinfo {title} {Plasmonic response of
  complex nanoparticle assemblies},\ }\href@noop {} {\bibfield  {journal}
  {\bibinfo  {journal} {Nano Lett.}\ }\textbf {\bibinfo {volume} {23}},\
  \bibinfo {pages} {3030} (\bibinfo {year} {2023})}\BibitemShut {NoStop}%
\bibitem [{\citenamefont {Ofosu}\ \emph {et~al.}(2026)\citenamefont {Ofosu},
  \citenamefont {Brackett}, \citenamefont {Conrad}, \citenamefont {Qian},
  \citenamefont {Lee}, \citenamefont {Kang}, \citenamefont {Choi},
  \citenamefont {Bessmertnaya}, \citenamefont {Oberlander}, \citenamefont
  {Green} \emph {et~al.}}]{ofosu2026universal}%
  \BibitemOpen
  \bibfield  {author} {\bibinfo {author} {\bibfnamefont {C.~K.}\ \bibnamefont
  {Ofosu}}, \bibinfo {author} {\bibfnamefont {W.~D.}\ \bibnamefont {Brackett}},
  \bibinfo {author} {\bibfnamefont {D.~L.}\ \bibnamefont {Conrad}}, \bibinfo
  {author} {\bibfnamefont {D.}~\bibnamefont {Qian}}, \bibinfo {author}
  {\bibfnamefont {T.-L.}\ \bibnamefont {Lee}}, \bibinfo {author} {\bibfnamefont
  {J.}~\bibnamefont {Kang}}, \bibinfo {author} {\bibfnamefont {J.}~\bibnamefont
  {Choi}}, \bibinfo {author} {\bibfnamefont {A.}~\bibnamefont {Bessmertnaya}},
  \bibinfo {author} {\bibfnamefont {J.~D.}\ \bibnamefont {Oberlander}},
  \bibinfo {author} {\bibfnamefont {A.~M.}\ \bibnamefont {Green}}, \emph
  {et~al.},\ }\bibfield  {title} {\bibinfo {title} {Universal progression of
  structure and dynamics in colloidal nanocrystal gels during salt-accelerated
  aging},\ }\href@noop {} {\bibfield  {journal} {\bibinfo  {journal} {Sci.
  Adv.}\ }\textbf {\bibinfo {volume} {12}},\ \bibinfo {pages} {eaec4820}
  (\bibinfo {year} {2026})}\BibitemShut {NoStop}%
\bibitem [{\citenamefont {Kang}\ \emph
  {et~al.}(2023{\natexlab{a}})\citenamefont {Kang}, \citenamefont {Sherman},
  \citenamefont {Crory}, \citenamefont {Conrad}, \citenamefont {Berry},
  \citenamefont {Roman}, \citenamefont {Anslyn}, \citenamefont {Truskett},\
  and\ \citenamefont {Milliron}}]{kang2023modular}%
  \BibitemOpen
  \bibfield  {author} {\bibinfo {author} {\bibfnamefont {J.}~\bibnamefont
  {Kang}}, \bibinfo {author} {\bibfnamefont {Z.~M.}\ \bibnamefont {Sherman}},
  \bibinfo {author} {\bibfnamefont {H.~S.~N.}\ \bibnamefont {Crory}}, \bibinfo
  {author} {\bibfnamefont {D.~L.}\ \bibnamefont {Conrad}}, \bibinfo {author}
  {\bibfnamefont {M.~W.}\ \bibnamefont {Berry}}, \bibinfo {author}
  {\bibfnamefont {B.~J.}\ \bibnamefont {Roman}}, \bibinfo {author}
  {\bibfnamefont {E.~V.}\ \bibnamefont {Anslyn}}, \bibinfo {author}
  {\bibfnamefont {T.~M.}\ \bibnamefont {Truskett}},\ and\ \bibinfo {author}
  {\bibfnamefont {D.~J.}\ \bibnamefont {Milliron}},\ }\bibfield  {title}
  {\bibinfo {title} {Modular mixing in plasmonic metal oxide nanocrystal gels
  with thermoreversible links},\ }\href@noop {} {\bibfield  {journal} {\bibinfo
   {journal} {J. Chem. Phys.}\ }\textbf {\bibinfo {volume} {158}},\ \bibinfo
  {pages} {024903} (\bibinfo {year} {2023}{\natexlab{a}})}\BibitemShut
  {NoStop}%
\bibitem [{\citenamefont {Kang}\ \emph
  {et~al.}(2023{\natexlab{b}})\citenamefont {Kang}, \citenamefont {Sherman},
  \citenamefont {Conrad}, \citenamefont {Crory}, \citenamefont {Dominguez},
  \citenamefont {Valenzuela}, \citenamefont {Anslyn}, \citenamefont
  {Truskett},\ and\ \citenamefont {Milliron}}]{kang2023structural}%
  \BibitemOpen
  \bibfield  {author} {\bibinfo {author} {\bibfnamefont {J.}~\bibnamefont
  {Kang}}, \bibinfo {author} {\bibfnamefont {Z.~M.}\ \bibnamefont {Sherman}},
  \bibinfo {author} {\bibfnamefont {D.~L.}\ \bibnamefont {Conrad}}, \bibinfo
  {author} {\bibfnamefont {H.~S.~N.}\ \bibnamefont {Crory}}, \bibinfo {author}
  {\bibfnamefont {M.~N.}\ \bibnamefont {Dominguez}}, \bibinfo {author}
  {\bibfnamefont {S.~A.}\ \bibnamefont {Valenzuela}}, \bibinfo {author}
  {\bibfnamefont {E.~V.}\ \bibnamefont {Anslyn}}, \bibinfo {author}
  {\bibfnamefont {T.~M.}\ \bibnamefont {Truskett}},\ and\ \bibinfo {author}
  {\bibfnamefont {D.~J.}\ \bibnamefont {Milliron}},\ }\bibfield  {title}
  {\bibinfo {title} {Structural control of plasmon resonance in molecularly
  linked metal oxide nanocrystal gel assemblies},\ }\href@noop {} {\bibfield
  {journal} {\bibinfo  {journal} {ACS Nano}\ }\textbf {\bibinfo {volume}
  {17}},\ \bibinfo {pages} {24218} (\bibinfo {year}
  {2023}{\natexlab{b}})}\BibitemShut {NoStop}%
\bibitem [{\citenamefont {Xu}\ \emph {et~al.}(2023)\citenamefont {Xu},
  \citenamefont {Choi}, \citenamefont {Nagella},\ and\ \citenamefont
  {Takatori}}]{xu2023dynamic}%
  \BibitemOpen
  \bibfield  {author} {\bibinfo {author} {\bibfnamefont {Y.}~\bibnamefont
  {Xu}}, \bibinfo {author} {\bibfnamefont {K.~H.}\ \bibnamefont {Choi}},
  \bibinfo {author} {\bibfnamefont {S.~G.}\ \bibnamefont {Nagella}},\ and\
  \bibinfo {author} {\bibfnamefont {S.~C.}\ \bibnamefont {Takatori}},\
  }\bibfield  {title} {\bibinfo {title} {Dynamic interfaces for contact-time
  control of colloidal interactions},\ }\href@noop {} {\bibfield  {journal}
  {\bibinfo  {journal} {Soft Matter}\ }\textbf {\bibinfo {volume} {19}},\
  \bibinfo {pages} {5692} (\bibinfo {year} {2023})}\BibitemShut {NoStop}%
\bibitem [{\citenamefont {Hallstrom}\ \emph {et~al.}(2026)\citenamefont
  {Hallstrom}, \citenamefont {Pan}, \citenamefont {Sia}, \citenamefont {Bae},
  \citenamefont {Qian}, \citenamefont {Qian}, \citenamefont {Liu},
  \citenamefont {Yao}, \citenamefont {Truskett}, \citenamefont {Milliron} \emph
  {et~al.}}]{hallstrom2026decoding}%
  \BibitemOpen
  \bibfield  {author} {\bibinfo {author} {\bibfnamefont {J.}~\bibnamefont
  {Hallstrom}}, \bibinfo {author} {\bibfnamefont {P.}~\bibnamefont {Pan}},
  \bibinfo {author} {\bibfnamefont {J.}~\bibnamefont {Sia}}, \bibinfo {author}
  {\bibfnamefont {S.}~\bibnamefont {Bae}}, \bibinfo {author} {\bibfnamefont
  {D.}~\bibnamefont {Qian}}, \bibinfo {author} {\bibfnamefont {C.}~\bibnamefont
  {Qian}}, \bibinfo {author} {\bibfnamefont {S.}~\bibnamefont {Liu}}, \bibinfo
  {author} {\bibfnamefont {L.}~\bibnamefont {Yao}}, \bibinfo {author}
  {\bibfnamefont {T.~M.}\ \bibnamefont {Truskett}}, \bibinfo {author}
  {\bibfnamefont {D.~J.}\ \bibnamefont {Milliron}}, \emph {et~al.},\ }\bibfield
   {title} {\bibinfo {title} {Decoding collective dynamics and complexity in
  nanoparticle assemblies using graph theory},\ }\href@noop {} {\bibfield
  {journal} {\bibinfo  {journal} {Science}\ }\textbf {\bibinfo {volume}
  {392}},\ \bibinfo {pages} {eaeb5134} (\bibinfo {year} {2026})}\BibitemShut
  {NoStop}%
\bibitem [{\citenamefont {Kwon}\ \emph {et~al.}(2026)\citenamefont {Kwon},
  \citenamefont {Singh}, \citenamefont {Sanchez-Gallegos}, \citenamefont
  {Perdomo-P{\'e}rez}, \citenamefont {Casta{\~n}eda-Priego},\ and\
  \citenamefont {Truskett}}]{kwon2026connectivity}%
  \BibitemOpen
  \bibfield  {author} {\bibinfo {author} {\bibfnamefont {T.}~\bibnamefont
  {Kwon}}, \bibinfo {author} {\bibfnamefont {M.}~\bibnamefont {Singh}},
  \bibinfo {author} {\bibfnamefont {J.~A.}\ \bibnamefont {Sanchez-Gallegos}},
  \bibinfo {author} {\bibfnamefont {R.}~\bibnamefont {Perdomo-P{\'e}rez}},
  \bibinfo {author} {\bibfnamefont {R.}~\bibnamefont {Casta{\~n}eda-Priego}},\
  and\ \bibinfo {author} {\bibfnamefont {T.~M.}\ \bibnamefont {Truskett}},\
  }\bibfield  {title} {\bibinfo {title} {Connectivity, rigidity, and gelation
  in linker-mediated colloidal networks},\ }\bibfield  {journal} {\bibinfo
  {journal} {ChemRxiv}\ }\href {https://doi.org/10.26434/chemrxiv.15004408/v2}
  {10.26434/chemrxiv.15004408/v2} (\bibinfo {year} {2026})\BibitemShut
  {NoStop}%
\bibitem [{\citenamefont {FitzSimons}\ \emph {et~al.}(2020)\citenamefont
  {FitzSimons}, \citenamefont {Oentoro}, \citenamefont {Shanbhag},
  \citenamefont {Anslyn},\ and\ \citenamefont
  {Rosales}}]{FitzSimons2020PreferentialAdditions}%
  \BibitemOpen
  \bibfield  {author} {\bibinfo {author} {\bibfnamefont {T.~M.}\ \bibnamefont
  {FitzSimons}}, \bibinfo {author} {\bibfnamefont {F.}~\bibnamefont {Oentoro}},
  \bibinfo {author} {\bibfnamefont {T.~V.}\ \bibnamefont {Shanbhag}}, \bibinfo
  {author} {\bibfnamefont {E.~V.}\ \bibnamefont {Anslyn}},\ and\ \bibinfo
  {author} {\bibfnamefont {A.~M.}\ \bibnamefont {Rosales}},\ }\bibfield
  {title} {\bibinfo {title} {Preferential control of forward reaction kinetics
  in hydrogels crosslinked with reversible conjugate additions},\ }\href
  {https://doi.org/10.1021/acs.macromol.0c00335} {\bibfield  {journal}
  {\bibinfo  {journal} {Macromolecules}\ }\textbf {\bibinfo {volume} {53}},\
  \bibinfo {pages} {3738} (\bibinfo {year} {2020})}\BibitemShut {NoStop}%
\bibitem [{\citenamefont {FitzSimons}\ \emph {et~al.}(2022)\citenamefont
  {FitzSimons}, \citenamefont {Anslyn},\ and\ \citenamefont
  {Rosales}}]{fitzsimons2022effect}%
  \BibitemOpen
  \bibfield  {author} {\bibinfo {author} {\bibfnamefont {T.~M.}\ \bibnamefont
  {FitzSimons}}, \bibinfo {author} {\bibfnamefont {E.~V.}\ \bibnamefont
  {Anslyn}},\ and\ \bibinfo {author} {\bibfnamefont {A.~M.}\ \bibnamefont
  {Rosales}},\ }\bibfield  {title} {\bibinfo {title} {{Effect of pH on the
  properties of hydrogels cross-linked via dynamic thia-Michael addition
  bonds}},\ }\href@noop {} {\bibfield  {journal} {\bibinfo  {journal} {ACS
  Polym. Au}\ }\textbf {\bibinfo {volume} {2}},\ \bibinfo {pages} {129}
  (\bibinfo {year} {2022})}\BibitemShut {NoStop}%
\bibitem [{\citenamefont {Crowell}\ \emph {et~al.}(2023)\citenamefont
  {Crowell}, \citenamefont {FitzSimons}, \citenamefont {Anslyn}, \citenamefont
  {Schultz},\ and\ \citenamefont {Rosales}}]{crowell2023shear}%
  \BibitemOpen
  \bibfield  {author} {\bibinfo {author} {\bibfnamefont {A.~D.}\ \bibnamefont
  {Crowell}}, \bibinfo {author} {\bibfnamefont {T.~M.}\ \bibnamefont
  {FitzSimons}}, \bibinfo {author} {\bibfnamefont {E.~V.}\ \bibnamefont
  {Anslyn}}, \bibinfo {author} {\bibfnamefont {K.~M.}\ \bibnamefont
  {Schultz}},\ and\ \bibinfo {author} {\bibfnamefont {A.~M.}\ \bibnamefont
  {Rosales}},\ }\bibfield  {title} {\bibinfo {title} {Shear thickening behavior
  in injectable tetra-{PEG} hydrogels cross-linked via dynamic thia-michael
  addition bonds},\ }\href@noop {} {\bibfield  {journal} {\bibinfo  {journal}
  {Macromolecules}\ }\textbf {\bibinfo {volume} {56}},\ \bibinfo {pages} {7795}
  (\bibinfo {year} {2023})}\BibitemShut {NoStop}%
\bibitem [{\citenamefont {Crowell}\ \emph {et~al.}(2026)\citenamefont
  {Crowell}, \citenamefont {Kang}, \citenamefont {Conrad}, \citenamefont
  {FitzSimons}, \citenamefont {Anslyn}, \citenamefont {Milliron},\ and\
  \citenamefont {Rosales}}]{crowell2026leveraging}%
  \BibitemOpen
  \bibfield  {author} {\bibinfo {author} {\bibfnamefont {A.~D.}\ \bibnamefont
  {Crowell}}, \bibinfo {author} {\bibfnamefont {J.}~\bibnamefont {Kang}},
  \bibinfo {author} {\bibfnamefont {D.~L.}\ \bibnamefont {Conrad}}, \bibinfo
  {author} {\bibfnamefont {T.~M.}\ \bibnamefont {FitzSimons}}, \bibinfo
  {author} {\bibfnamefont {E.~V.}\ \bibnamefont {Anslyn}}, \bibinfo {author}
  {\bibfnamefont {D.~J.}\ \bibnamefont {Milliron}},\ and\ \bibinfo {author}
  {\bibfnamefont {A.~M.}\ \bibnamefont {Rosales}},\ }\bibfield  {title}
  {\bibinfo {title} {Leveraging bond dissociation kinetics to tune
  shear-thickening behavior in dynamic covalent tetra-{PEG} hydrogels},\
  }\href@noop {} {\bibfield  {journal} {\bibinfo  {journal} {Sci. Adv.}\
  }\textbf {\bibinfo {volume} {12}},\ \bibinfo {pages} {eadz9563} (\bibinfo
  {year} {2026})}\BibitemShut {NoStop}%
\bibitem [{\citenamefont {Oyarz{\'{u}}n}\ and\ \citenamefont
  {Mognetti}(2018)}]{Oyarzun2018EfficientNanoparticles}%
  \BibitemOpen
  \bibfield  {author} {\bibinfo {author} {\bibfnamefont {B.}~\bibnamefont
  {Oyarz{\'{u}}n}}\ and\ \bibinfo {author} {\bibfnamefont {B.~M.}\ \bibnamefont
  {Mognetti}},\ }\bibfield  {title} {\bibinfo {title} {{Efficient sampling of
  reversible cross-linking polymers: Self-assembly of single-chain polymeric
  nanoparticles}},\ }\bibfield  {journal} {\bibinfo  {journal} {J. Chem.
  Phys.}\ }\textbf {\bibinfo {volume} {148}},\ \href
  {https://doi.org/10.1063/1.5020158} {10.1063/1.5020158} (\bibinfo {year}
  {2018})\BibitemShut {NoStop}%
\bibitem [{\citenamefont {Zhang}\ and\ \citenamefont
  {DuBay}(2021)}]{zhang2021sequence}%
  \BibitemOpen
  \bibfield  {author} {\bibinfo {author} {\bibfnamefont {Z.}~\bibnamefont
  {Zhang}}\ and\ \bibinfo {author} {\bibfnamefont {K.~H.}\ \bibnamefont
  {DuBay}},\ }\bibfield  {title} {\bibinfo {title} {The sequence of a
  step-growth copolymer can be influenced by its own persistence length},\
  }\href@noop {} {\bibfield  {journal} {\bibinfo  {journal} {J. Phys. Chem. B}\
  }\textbf {\bibinfo {volume} {125}},\ \bibinfo {pages} {3426} (\bibinfo {year}
  {2021})}\BibitemShut {NoStop}%
\bibitem [{\citenamefont {Dai}\ \emph {et~al.}(2026)\citenamefont {Dai},
  \citenamefont {Qiu}, \citenamefont {Leon~Hernandez}, \citenamefont {Chan},
  \citenamefont {Mejia}, \citenamefont {Nemeria},\ and\ \citenamefont
  {Kinz-Thompson}}]{dai2026modular}%
  \BibitemOpen
  \bibfield  {author} {\bibinfo {author} {\bibfnamefont {T.}~\bibnamefont
  {Dai}}, \bibinfo {author} {\bibfnamefont {Y.}~\bibnamefont {Qiu}}, \bibinfo
  {author} {\bibfnamefont {K.~M.}\ \bibnamefont {Leon~Hernandez}}, \bibinfo
  {author} {\bibfnamefont {W.}~\bibnamefont {Chan}}, \bibinfo {author}
  {\bibfnamefont {K.}~\bibnamefont {Mejia}}, \bibinfo {author} {\bibfnamefont
  {N.}~\bibnamefont {Nemeria}},\ and\ \bibinfo {author} {\bibfnamefont {C.~D.}\
  \bibnamefont {Kinz-Thompson}},\ }\bibfield  {title} {\bibinfo {title}
  {Modular assembly of dynamic polymer networks from heteroaffinity cross-links
  to multivalent proteins},\ }\href@noop {} {\bibfield  {journal} {\bibinfo
  {journal} {Angew. Chem.}\ }\textbf {\bibinfo {volume} {138}},\ \bibinfo
  {pages} {e26058} (\bibinfo {year} {2026})}\BibitemShut {NoStop}%
\bibitem [{\citenamefont {Angioletti-Uberti}\ \emph {et~al.}(2014)\citenamefont
  {Angioletti-Uberti}, \citenamefont {Varilly}, \citenamefont {Mognetti},\ and\
  \citenamefont {Frenkel}}]{angioletti2014mobile}%
  \BibitemOpen
  \bibfield  {author} {\bibinfo {author} {\bibfnamefont {S.}~\bibnamefont
  {Angioletti-Uberti}}, \bibinfo {author} {\bibfnamefont {P.}~\bibnamefont
  {Varilly}}, \bibinfo {author} {\bibfnamefont {B.~M.}\ \bibnamefont
  {Mognetti}},\ and\ \bibinfo {author} {\bibfnamefont {D.}~\bibnamefont
  {Frenkel}},\ }\bibfield  {title} {\bibinfo {title} {Mobile linkers on
  {DNA}-coated colloids: valency without patches},\ }\href@noop {} {\bibfield
  {journal} {\bibinfo  {journal} {Phys. Rev. Lett.}\ }\textbf {\bibinfo
  {volume} {113}},\ \bibinfo {pages} {128303} (\bibinfo {year}
  {2014})}\BibitemShut {NoStop}%
\bibitem [{\citenamefont {Bachmann}\ \emph {et~al.}(2016)\citenamefont
  {Bachmann}, \citenamefont {Kotar}, \citenamefont {Parolini}, \citenamefont
  {{\v{S}}ari{\'c}}, \citenamefont {Cicuta}, \citenamefont {Di~Michele},\ and\
  \citenamefont {Mognetti}}]{bachmann2016melting}%
  \BibitemOpen
  \bibfield  {author} {\bibinfo {author} {\bibfnamefont {S.~J.}\ \bibnamefont
  {Bachmann}}, \bibinfo {author} {\bibfnamefont {J.}~\bibnamefont {Kotar}},
  \bibinfo {author} {\bibfnamefont {L.}~\bibnamefont {Parolini}}, \bibinfo
  {author} {\bibfnamefont {A.}~\bibnamefont {{\v{S}}ari{\'c}}}, \bibinfo
  {author} {\bibfnamefont {P.}~\bibnamefont {Cicuta}}, \bibinfo {author}
  {\bibfnamefont {L.}~\bibnamefont {Di~Michele}},\ and\ \bibinfo {author}
  {\bibfnamefont {B.~M.}\ \bibnamefont {Mognetti}},\ }\bibfield  {title}
  {\bibinfo {title} {Melting transition in lipid vesicles functionalised by
  mobile {DNA} linkers},\ }\href@noop {} {\bibfield  {journal} {\bibinfo
  {journal} {Soft Matter}\ }\textbf {\bibinfo {volume} {12}},\ \bibinfo {pages}
  {7804} (\bibinfo {year} {2016})}\BibitemShut {NoStop}%
\bibitem [{\citenamefont {McMullen}\ \emph {et~al.}(2018)\citenamefont
  {McMullen}, \citenamefont {Holmes-Cerfon}, \citenamefont {Sciortino},
  \citenamefont {Grosberg},\ and\ \citenamefont {Brujic}}]{mcmullen2018freely}%
  \BibitemOpen
  \bibfield  {author} {\bibinfo {author} {\bibfnamefont {A.}~\bibnamefont
  {McMullen}}, \bibinfo {author} {\bibfnamefont {M.}~\bibnamefont
  {Holmes-Cerfon}}, \bibinfo {author} {\bibfnamefont {F.}~\bibnamefont
  {Sciortino}}, \bibinfo {author} {\bibfnamefont {A.~Y.}\ \bibnamefont
  {Grosberg}},\ and\ \bibinfo {author} {\bibfnamefont {J.}~\bibnamefont
  {Brujic}},\ }\bibfield  {title} {\bibinfo {title} {Freely jointed polymers
  made of droplets},\ }\href@noop {} {\bibfield  {journal} {\bibinfo  {journal}
  {Phys. Rev. Lett.}\ }\textbf {\bibinfo {volume} {121}},\ \bibinfo {pages}
  {138002} (\bibinfo {year} {2018})}\BibitemShut {NoStop}%
\bibitem [{\citenamefont {Judd}\ \emph {et~al.}(2025)\citenamefont {Judd},
  \citenamefont {McMullen}, \citenamefont {Hilgenfeldt},\ and\ \citenamefont
  {Brujic}}]{judd2025statistical}%
  \BibitemOpen
  \bibfield  {author} {\bibinfo {author} {\bibfnamefont {N.}~\bibnamefont
  {Judd}}, \bibinfo {author} {\bibfnamefont {A.}~\bibnamefont {McMullen}},
  \bibinfo {author} {\bibfnamefont {S.}~\bibnamefont {Hilgenfeldt}},\ and\
  \bibinfo {author} {\bibfnamefont {J.}~\bibnamefont {Brujic}},\ }\bibfield
  {title} {\bibinfo {title} {Statistical mechanics approach to {DNA}-driven
  droplet deformation and adhesion},\ }\href@noop {} {\bibfield  {journal}
  {\bibinfo  {journal} {Phys. Rev. Lett.}\ }\textbf {\bibinfo {volume} {134}},\
  \bibinfo {pages} {058202} (\bibinfo {year} {2025})}\BibitemShut {NoStop}%
\bibitem [{\citenamefont {Scheutz}\ \emph {et~al.}(2019)\citenamefont
  {Scheutz}, \citenamefont {Lessard}, \citenamefont {Sims},\ and\ \citenamefont
  {Sumerlin}}]{scheutz2019adaptable}%
  \BibitemOpen
  \bibfield  {author} {\bibinfo {author} {\bibfnamefont {G.~M.}\ \bibnamefont
  {Scheutz}}, \bibinfo {author} {\bibfnamefont {J.~J.}\ \bibnamefont
  {Lessard}}, \bibinfo {author} {\bibfnamefont {M.~B.}\ \bibnamefont {Sims}},\
  and\ \bibinfo {author} {\bibfnamefont {B.~S.}\ \bibnamefont {Sumerlin}},\
  }\bibfield  {title} {\bibinfo {title} {Adaptable crosslinks in polymeric
  materials: resolving the intersection of thermoplastics and thermosets},\
  }\href@noop {} {\bibfield  {journal} {\bibinfo  {journal} {J. Am. Chem.
  Soc.}\ }\textbf {\bibinfo {volume} {141}},\ \bibinfo {pages} {16181}
  (\bibinfo {year} {2019})}\BibitemShut {NoStop}%
\bibitem [{\citenamefont {Reuther}\ \emph {et~al.}(2019)\citenamefont
  {Reuther}, \citenamefont {Dahlhauser},\ and\ \citenamefont
  {Anslyn}}]{reuther2019tunable}%
  \BibitemOpen
  \bibfield  {author} {\bibinfo {author} {\bibfnamefont {J.~F.}\ \bibnamefont
  {Reuther}}, \bibinfo {author} {\bibfnamefont {S.~D.}\ \bibnamefont
  {Dahlhauser}},\ and\ \bibinfo {author} {\bibfnamefont {E.~V.}\ \bibnamefont
  {Anslyn}},\ }\bibfield  {title} {\bibinfo {title} {Tunable orthogonal
  reversible covalent {(TORC)} bonds: dynamic chemical control over molecular
  assembly},\ }\href@noop {} {\bibfield  {journal} {\bibinfo  {journal} {Angew.
  Chem. Int. Ed.}\ }\textbf {\bibinfo {volume} {58}},\ \bibinfo {pages} {74}
  (\bibinfo {year} {2019})}\BibitemShut {NoStop}%
\bibitem [{\citenamefont {Cudjoe}\ \emph {et~al.}(2018)\citenamefont {Cudjoe},
  \citenamefont {Herbert},\ and\ \citenamefont {Rowan}}]{cudjoe2018strong}%
  \BibitemOpen
  \bibfield  {author} {\bibinfo {author} {\bibfnamefont {E.}~\bibnamefont
  {Cudjoe}}, \bibinfo {author} {\bibfnamefont {K.~M.}\ \bibnamefont
  {Herbert}},\ and\ \bibinfo {author} {\bibfnamefont {S.~J.}\ \bibnamefont
  {Rowan}},\ }\bibfield  {title} {\bibinfo {title} {Strong, rebondable, dynamic
  cross-linked cellulose nanocrystal polymer nanocomposite adhesives},\
  }\href@noop {} {\bibfield  {journal} {\bibinfo  {journal} {ACS Appl. Mater.
  Interfaces}\ }\textbf {\bibinfo {volume} {10}},\ \bibinfo {pages} {30723}
  (\bibinfo {year} {2018})}\BibitemShut {NoStop}%
\bibitem [{\citenamefont {Zhang}\ \emph {et~al.}(2022)\citenamefont {Zhang},
  \citenamefont {Kang}, \citenamefont {Accardo},\ and\ \citenamefont
  {Kalow}}]{Zhang2022Structure-Reactivity-PropertyNetworks}%
  \BibitemOpen
  \bibfield  {author} {\bibinfo {author} {\bibfnamefont {V.}~\bibnamefont
  {Zhang}}, \bibinfo {author} {\bibfnamefont {B.}~\bibnamefont {Kang}},
  \bibinfo {author} {\bibfnamefont {J.~V.}\ \bibnamefont {Accardo}},\ and\
  \bibinfo {author} {\bibfnamefont {J.~A.}\ \bibnamefont {Kalow}},\ }\bibfield
  {title} {\bibinfo {title} {Structure–reactivity–property relationships in
  covalent adaptable networks},\ }\href {https://doi.org/10.1021/jacs.2c08104}
  {\bibfield  {journal} {\bibinfo  {journal} {J. Am. Chem. Soc.}\ }\textbf
  {\bibinfo {volume} {144}},\ \bibinfo {pages} {22358} (\bibinfo {year}
  {2022})}\BibitemShut {NoStop}%
\bibitem [{\citenamefont {Anderson}\ \emph {et~al.}(2020)\citenamefont
  {Anderson}, \citenamefont {Glaser},\ and\ \citenamefont
  {Glotzer}}]{Anderson2020HOOMD-blue:Simulations}%
  \BibitemOpen
  \bibfield  {author} {\bibinfo {author} {\bibfnamefont {J.~A.}\ \bibnamefont
  {Anderson}}, \bibinfo {author} {\bibfnamefont {J.}~\bibnamefont {Glaser}},\
  and\ \bibinfo {author} {\bibfnamefont {S.~C.}\ \bibnamefont {Glotzer}},\
  }\bibfield  {title} {\bibinfo {title} {{HOOMD-blue: A Python package for
  high-performance molecular dynamics and hard particle Monte Carlo
  simulations}},\ }\href {https://doi.org/10.1016/j.commatsci.2019.109363}
  {\bibfield  {journal} {\bibinfo  {journal} {Comput. Mater. Sci.}\ }\textbf
  {\bibinfo {volume} {173}},\ \bibinfo {pages} {109363} (\bibinfo {year}
  {2020})}\BibitemShut {NoStop}%
\bibitem [{\citenamefont {Thomas}\ \emph {et~al.}(2018)\citenamefont {Thomas},
  \citenamefont {Alberts}, \citenamefont {Henry}, \citenamefont {Estridge},\
  and\ \citenamefont {Jankowski}}]{Thomas2018RoutineDynamics}%
  \BibitemOpen
  \bibfield  {author} {\bibinfo {author} {\bibfnamefont {S.}~\bibnamefont
  {Thomas}}, \bibinfo {author} {\bibfnamefont {M.}~\bibnamefont {Alberts}},
  \bibinfo {author} {\bibfnamefont {M.~M.}\ \bibnamefont {Henry}}, \bibinfo
  {author} {\bibfnamefont {C.~E.}\ \bibnamefont {Estridge}},\ and\ \bibinfo
  {author} {\bibfnamefont {E.}~\bibnamefont {Jankowski}},\ }\bibfield  {title}
  {\bibinfo {title} {{Routine million-particle simulations of epoxy curing with
  dissipative particle dynamics}},\ }\href
  {https://doi.org/10.1142/S0219633618400059} {\bibfield  {journal} {\bibinfo
  {journal} {J. Theor. Comput. Chem.}\ }\textbf {\bibinfo {volume} {17}},\
  \bibinfo {pages} {1840005} (\bibinfo {year} {2018})}\BibitemShut {NoStop}%
\bibitem [{\citenamefont {Marbach}\ and\ \citenamefont
  {Miles}(2023)}]{Marbach2023Coarse-grainedLinkers}%
  \BibitemOpen
  \bibfield  {author} {\bibinfo {author} {\bibfnamefont {S.}~\bibnamefont
  {Marbach}}\ and\ \bibinfo {author} {\bibfnamefont {C.~E.}\ \bibnamefont
  {Miles}},\ }\bibfield  {title} {\bibinfo {title} {{Coarse-grained dynamics of
  transiently bound fast linkers}},\ }\bibfield  {journal} {\bibinfo  {journal}
  {J. Comp. Phys.}\ }\textbf {\bibinfo {volume} {158}},\ \href
  {https://doi.org/10.1063/5.0139036} {10.1063/5.0139036} (\bibinfo {year}
  {2023})\BibitemShut {NoStop}%
\bibitem [{\citenamefont {Metropolis}\ \emph {et~al.}(1953)\citenamefont
  {Metropolis}, \citenamefont {Rosenbluth}, \citenamefont {Rosenbluth},
  \citenamefont {Teller},\ and\ \citenamefont
  {Teller}}]{metropolis1953equation}%
  \BibitemOpen
  \bibfield  {author} {\bibinfo {author} {\bibfnamefont {N.}~\bibnamefont
  {Metropolis}}, \bibinfo {author} {\bibfnamefont {A.~W.}\ \bibnamefont
  {Rosenbluth}}, \bibinfo {author} {\bibfnamefont {M.~N.}\ \bibnamefont
  {Rosenbluth}}, \bibinfo {author} {\bibfnamefont {A.~H.}\ \bibnamefont
  {Teller}},\ and\ \bibinfo {author} {\bibfnamefont {E.}~\bibnamefont
  {Teller}},\ }\bibfield  {title} {\bibinfo {title} {Equation of state
  calculations by fast computing machines},\ }\href@noop {} {\bibfield
  {journal} {\bibinfo  {journal} {J. Chem. Phys.}\ }\textbf {\bibinfo {volume}
  {21}},\ \bibinfo {pages} {1087} (\bibinfo {year} {1953})}\BibitemShut
  {NoStop}%
\bibitem [{\citenamefont {Bell}(1978)}]{bell1978models}%
  \BibitemOpen
  \bibfield  {author} {\bibinfo {author} {\bibfnamefont {G.~I.}\ \bibnamefont
  {Bell}},\ }\bibfield  {title} {\bibinfo {title} {Models for the specific
  adhesion of cells to cells},\ }\href {https://doi.org/10.1126/science.347575}
  {\bibfield  {journal} {\bibinfo  {journal} {Science}\ }\textbf {\bibinfo
  {volume} {200}},\ \bibinfo {pages} {618} (\bibinfo {year}
  {1978})}\BibitemShut {NoStop}%
\bibitem [{\citenamefont {Salmon}\ \emph {et~al.}(2011)\citenamefont {Salmon},
  \citenamefont {Moraes}, \citenamefont {Dror},\ and\ \citenamefont
  {Shaw}}]{salmon2011parallel}%
  \BibitemOpen
  \bibfield  {author} {\bibinfo {author} {\bibfnamefont {J.~K.}\ \bibnamefont
  {Salmon}}, \bibinfo {author} {\bibfnamefont {M.~A.}\ \bibnamefont {Moraes}},
  \bibinfo {author} {\bibfnamefont {R.~O.}\ \bibnamefont {Dror}},\ and\
  \bibinfo {author} {\bibfnamefont {D.~E.}\ \bibnamefont {Shaw}},\ }\bibfield
  {title} {\bibinfo {title} {Parallel random numbers: as easy as 1, 2, 3},\
  }in\ \href {https://doi.org/10.1145/2063384.2063405} {\emph {\bibinfo
  {booktitle} {Proceedings of 2011 International Conference for High
  Performance Computing, Networking, Storage and Analysis}}}\ (\bibinfo
  {publisher} {ACM},\ \bibinfo {year} {2011})\ pp.\ \bibinfo {pages}
  {1--12}\BibitemShut {NoStop}%
\bibitem [{\citenamefont {Warner~Jr}(1972)}]{warner1972kinetic}%
  \BibitemOpen
  \bibfield  {author} {\bibinfo {author} {\bibfnamefont {H.~R.}\ \bibnamefont
  {Warner~Jr}},\ }\bibfield  {title} {\bibinfo {title} {Kinetic theory and
  rheology of dilute suspensions of finitely extendible dumbbells},\
  }\href@noop {} {\bibfield  {journal} {\bibinfo  {journal} {Ind. Eng. Chem.
  Fundam.}\ }\textbf {\bibinfo {volume} {11}},\ \bibinfo {pages} {379}
  (\bibinfo {year} {1972})}\BibitemShut {NoStop}%
\bibitem [{\citenamefont {Weeks}\ \emph {et~al.}(1971)\citenamefont {Weeks},
  \citenamefont {Chandler},\ and\ \citenamefont {Andersen}}]{weeks1971role}%
  \BibitemOpen
  \bibfield  {author} {\bibinfo {author} {\bibfnamefont {J.~D.}\ \bibnamefont
  {Weeks}}, \bibinfo {author} {\bibfnamefont {D.}~\bibnamefont {Chandler}},\
  and\ \bibinfo {author} {\bibfnamefont {H.~C.}\ \bibnamefont {Andersen}},\
  }\bibfield  {title} {\bibinfo {title} {Role of repulsive forces in
  determining the equilibrium structure of simple liquids},\ }\href@noop {}
  {\bibfield  {journal} {\bibinfo  {journal} {J. Chem. Phys.}\ }\textbf
  {\bibinfo {volume} {54}},\ \bibinfo {pages} {5237} (\bibinfo {year}
  {1971})}\BibitemShut {NoStop}%
\end{thebibliography}%

\end{document}